\documentclass[aps,prd,10pt,nofootinbib,OliveGeqsecnum,showpacs,showkeys,superscriptaddress,preprintnumbers,altaffilletter,floatfix,twocolumn]{revtex4-2}

\usepackage{graphicx,caption,subcaption,bm,color,hyperref}
\usepackage{amsmath,amssymb}
\usepackage{mathtools} 
\usepackage{dsfont} 
\usepackage{multirow} 
\usepackage{float} 
\usepackage{mathrsfs} 
\usepackage{tensor} 
\usepackage{booktabs}
\usepackage{ragged2e}
\usepackage{needspace}   
\usepackage{placeins}    
\usepackage{afterpage}

\hypersetup{colorlinks, linkcolor={blue}, citecolor={red}, urlcolor={teal}}

\definecolor{amaranth}{rgb}{0.9, 0.17, 0.31}
\definecolor{palatinateblue}{rgb}{0.15, 0.23, 0.89}
\definecolor{brightpink}{rgb}{1.0, 0.0, 0.5}

\definecolor{forestgreen}{rgb}{0.13, 0.55, 0.13}
\definecolor{goldenrod}{rgb}{0.85, 0.65, 0.13}
\definecolor{darkviolet}{rgb}{0.58, 0.0, 0.83}
\definecolor{turquoise}{rgb}{0.0, 0.78, 0.8}

\hypersetup{ 
    linktoc=all,
    colorlinks, 
    linkcolor={palatinateblue},
    citecolor={brightpink}, 
    urlcolor={amaranth}
}

\graphicspath{{images/}}

\newcommand{\be}{\begin{equation}}
\newcommand{\ee}{\end{equation}}
\newcommand{\ba}{\begin{eqnarray}}
\newcommand{\ea}{\end{eqnarray}}

\newcommand{\mH}{\mathcal{H}}

\newcommand{\nn}{\nonumber} 
\newcommand{\C}{\mathcal{C}}

\begin{document}

\title{Addressing $H_0$ and $S_8$ tensions within $f(Q)$ cosmology}

\author{Carlos G. Boiza}
\email{carlos.garciab@ehu.eus}
\affiliation{Department of Physics \& EHU Quantum Center, University of the 
Basque Country UPV/EHU, P.O. Box 644, 48080 Bilbao, Spain}
\author{Maria Petronikolou}

\email{petronikoloum@gmail.com}
\affiliation{Department of Physics, National Technical University of Athens, 
Zografou
Campus GR 157 73, Athens, Greece}
\affiliation{National Observatory of Athens, Lofos Nymfon, 11852 Athens, 
Greece}

\author{Mariam Bouhmadi-López}
\email{mariam.bouhmadi@ehu.eus}
\affiliation{IKERBASQUE, Basque Foundation for Science, 48011, Bilbao, Spain}
\affiliation{Department of Physics  \& EHU Quantum Center, University of the 
Basque Country UPV/EHU, P.O. Box 644, 48080 Bilbao, Spain}

\author{Emmanuel N. Saridakis}
\email{msaridak@noa.gr}
\affiliation{National Observatory of Athens, Lofos Nymfon, 11852 Athens, 
Greece}
\affiliation{CAS Key Laboratory for Researches in Galaxies and Cosmology, 
Department of Astronomy, University of Science and Technology of China, Hefei, 
Anhui 230026, P.R. China}
\affiliation{Departamento de Matem\'{a}ticas, Universidad Cat\'{o}lica del 
Norte, Avda. Angamos 0610, Casilla 1280 Antofagasta, Chile}

\begin{abstract} 

We investigate the viability of $f(Q)$ gravity as an alternative framework to address the $H_0$ and $S_8$ tensions in cosmology. Focusing on three representative $f(Q)$ models, we perform a comprehensive Bayesian analysis using a combination of cosmological observations, including cosmic chronometers, Type Ia supernovae, gamma-ray bursts, baryon acoustic oscillations, and CMB distance priors. Our results demonstrate that most of these models can yield higher values of $H_0$ than those predicted by $\Lambda$CDM, offering a partial alleviation of the tension. In addition, one model satisfies the condition $G_{\mathrm{eff}} < G$, making it a promising candidate for addressing the $S_8$ tension. However, these improvements are accompanied by mild internal inconsistencies between different subsets of data, which limit the overall statistical preference relative to $\Lambda$CDM. Despite this, $f(Q)$ gravity remains a promising and flexible framework for late-time cosmology, and our results motivate further exploration of extended or hybrid models that may reconcile all observational constraints.

\end{abstract}

\maketitle

\renewcommand{\tocname}{Index}


\section{Introduction}\label{intro}

The Standard Model of Cosmology, known as $\Lambda$-Cold Dark Matter 
($\Lambda$CDM) combined with inflation within the framework of general 
relativity, has proven to be highly effective in describing the evolution of 
the 
universe, both at the background level and in terms of perturbations 
\cite{Peebles:2002gy}. Nevertheless, the possibility that late-time acceleration 
may have a dynamical origin has motivated the development of numerous extensions 
and modifications.
These extensions can generally be categorised into two main classes. The first 
class retains general relativity as the foundational gravitational framework 
but 
incorporates additional components, such as dark energy sectors 
\cite{Copeland:2006wr,Cai:2009zp}. The second class involves constructing 
modified gravity theories that include general relativity as a special case, 
while generally introducing additional degrees of freedom capable of driving 
the 
acceleration of the universe 
\cite{CANTATA:2021asi,Capozziello:2011et,Cai:2015emx}.

There are numerous approaches to constructing modifications of gravity. In the 
simplest cases, one starts with the Einstein-Hilbert Lagrangian and introduces 
additional terms, leading to theories such as $f(R)$ gravity 
\cite{Starobinsky:1980te}, $f(G)$ gravity \cite{Nojiri:2005jg}, $f(P)$ gravity 
\cite{Erices:2019mkd}, Lovelock gravity \cite{Lovelock:1971yv}, and 
Horndeski/Galileon scalar-tensor theories 
\cite{Horndeski:1974wa,Deffayet:2009wt}, among others.
Alternatively, one can begin with the torsion-based formulation of gravity and 
modify it, giving rise to theories such as $f(T)$ gravity 
\cite{Bengochea:2008gz,Cai:2015emx}, $f(T,T_{G})$ gravity 
\cite{Kofinas:2014owa}, $f(T,B)$ gravity \cite{Bahamonde:2015zma}, 
scalar-torsion theories \cite{Geng:2011aj}, and more.

Another distinct class of gravitational modifications is based on the 
equivalent 
formulation of gravity using non-metricity. This approach, initiated by  
\cite{Nester:1998mp,BeltranJimenez:2017tkd}, relies on an affine connection 
characterised by vanishing curvature and torsion but metric incompatibility. 
Recently, this framework has been extended to $f(Q)$ gravity 
\cite{BeltranJimenez:2017tkd}.
The $f(Q)$ gravity theory includes general relativity as a specific limit and 
benefits from possessing second-order field equations. These features have 
sparked significant interest in its cosmological applications within the 
scientific literature 
\cite{BeltranJimenez:2019tme,Lazkoz:2019sjl,Anagnostopoulos:2021ydo,Lu:2019hra,
Mandal:2020buf,Ayuso:2020dcu,Bajardi:2020fxh,
Frusciante:2021sio,
Atayde:2021pgb,
Ferreira:2022jcd,Capozziello:2022wgl,Gadbail:2022jco,
Sarmah:2023oum,
Khyllep:2021pcu,Barros:2020bgg,
De:2022shr,
 Solanki:2022rwu,
Solanki:2022ccf,
Beh:2021wva,
Lymperis:2022oyo,
DAmbrosio:2021zpm,Li:2021mdp,Dimakis:2021gby,
Hohmann:2021ast, 
Kar:2021juu,
Wang:2021zaz,Quiros:2021eju,Mandal:2021bpd,Albuquerque:2022eac,
Papagiannopoulos:2022ohv,Anagnostopoulos:2022gej,Arora:2022mlo,
Pati:2022dwl, 
Maurya:2022wwa,Capozziello:2022tvv,Dimakis:2022wkj,DAgostino:2022tdk,
Narawade:2022cgb,Emtsova:2022uij,Bahamonde:2022cmz,
Narawade:2023rip,
Ferreira:2023tat,Shabani:2023xfn,Sokoliuk:2023ccw,Shaikh:2023tii,Jan:2023djj,Dimakis:2023uib,
Koussour:2023rly,Najera:2023wcw,Atayde:2023aoj,Shabani:2024ler,
Yang:2024tkw,Wang:2024dkn,Wang:2024eai,Vasquez:2025hrz,
Nashed:2025usa,ElOuardi:2025okl}. 

In recent years, an additional motivation for extending or modifying the 
concordance cosmology has emerged, driven by the need to address existing 
tensions, such as the $H_0$ and $S_8$ discrepancies \cite{Perivolaropoulos:2021jda,Abdalla:2022yfr} (see \cite{CosmoVerse:2025txj} for a review). The $H_0$ tension stems from 
the fact that the Planck collaboration's estimation of the present-day cosmic 
expansion rate is $H_0 = (67.27 \pm 0.60)$ km/s/Mpc \cite{Planck:2018vyg}, 
which exhibits a tension of approximately $4.4\sigma$ with the direct 
measurement obtained by the 2019 SH0ES collaboration (R19), namely $H_0 = 
(74.03 \pm 1.42)$ km/s/Mpc. The latter was derived using Hubble Space Telescope 
observations of 70 long-period Cepheids in the Large Magellanic Cloud 
\cite{Riess:2019cxk}. Furthermore, combining this with gravitational lensing 
and time-delay data increases the deviation to $5.3\sigma$ 
\cite{H0LiCOW:2019pvv}.
Similarly, the $S_8$ tension is associated with the parameter that 
quantifies matter clustering within spheres of radius $8$h$^{-1}$ Mpc. It 
arises from a possible discrepancy between estimates based on the Cosmic 
Microwave Background (CMB) \cite{Planck:2018vyg} and those obtained from 
SDSS/BOSS measurements \cite{Zarrouk:2018vwy, Alam:2016hwk, Ata:2017dya}.
If these tensions are not due to unknown systematics - which, at least in the 
case of the $H_0$ tension, appears increasingly unlikely - then it becomes 
necessary to explore extensions to the standard cosmological framework to 
address them effectively.

In this study, we focus on addressing the $H_0$ and $S_8$ tensions within the framework of $f(Q)$ gravity. We analyse three different models and we confront them with observational data from both background and large-scale structure probes. While we show that a partial alleviation of the $H_0$ tension is achievable for most models, it comes at a cost, namely, the emergence of internal tensions between different dataset combinations, which ultimately degrade the global fit compared to $\Lambda$CDM. The structure of the paper is as follows: In Section \ref{fQmodel}, we review $f(Q)$ gravity and $f(Q)$ cosmology, providing the equations at both background and perturbative levels, and present the specific models under study. In Section \ref{Observations}, we describe the datasets, outline our methodology, discuss the statistical tools used for model comparison, and present our results. Finally, we summarise our findings and conclude in Section \ref{conclusions}.

\section{\texorpdfstring{$f(Q)$} {f(Q)} gravity and cosmology}
\label{fQmodel}

In this section, we provide a concise overview of $f(Q)$ gravity and examine 
its 
application within a cosmological framework.

\subsection{\texorpdfstring{$f(Q)$} {f(Q)} gravity}

We begin by introducing the fundamental tools and framework of modified gravity 
theories based on non-metricity. The general affine connection 
$\Gamma^{\alpha}_{\mu\nu}$ can be expressed as a decomposition: 
\begin{equation}
 \Gamma^{\alpha}_{\mu\nu}=\hat{\Gamma}^{\alpha}_{\mu\nu}+
K^{\alpha}_{\,\,\mu\nu}+L^{\alpha}_{\,\,\mu\nu},
\end{equation}
where 
$ \hat{\Gamma}^{\alpha}_{\mu\nu}$
 is the Levi-Civita connection,
\begin{equation}
K^{\alpha}_{\,\,\mu\nu}=\frac{1}{2}T^{\alpha}_{\,\,\mu\nu}+T^{\,\,\,\alpha}_{
(\mu\,\,\,\nu)}
\end{equation}
is the contortion tensor derived from the   torsion 
tensor $T^{\alpha}_{\,\,\mu\nu}$, and
\begin{equation}
L^{\alpha}_{\,\,\mu\nu}=\frac{1}{2}Q^{\alpha}_{\,\,\mu\nu}-Q^{\,\,\,\alpha}_{
(\mu\,\,\,\nu)}
\end{equation}
is the disformation tensor  which arises due to the presence of non-metricity 
\begin{equation}
    Q_{\alpha\mu\nu}\equiv\nabla_\alpha g_{\mu\nu},
\end{equation}
with $g_{\mu\nu}$ the metric 
(Greek indices are used throughout to 
represent the coordinate space).
Using the general affine connection, the torsion and curvature tensors can be 
written as follows:
 \begin{eqnarray}
\label{Tortnsor}
&&\!\!\!\!\!\!\!\!\!\!\!
T^{\lambda}{}_{\mu\nu}\equiv 
\Gamma^{\lambda}{}_{\mu\nu}-\Gamma^{\lambda}{}_{\nu\mu}\,\nonumber\\
&&\!\!\!\!\!\!\!\!\!\!\!
R^{\sigma}{}_{\rho\mu\nu} \equiv \partial_{\mu} \Gamma^{\sigma}{}_{\nu\rho} - 
\partial_{\nu} \Gamma^{\sigma}{}_{\mu\rho} + \Gamma^{\alpha}{}_{\nu\rho} 
\Gamma^{\sigma}{}_{\mu\alpha} - \Gamma^{\alpha}{}_{\mu\rho} 
\Gamma^{\sigma}{}_{\nu\alpha}, \ 
\label{Rietsor}
\end{eqnarray}
while the non-metricity tensor can be expressed as
\begin{equation}
\label{NonMetrensor}
Q_{\rho \mu \nu} \equiv \nabla_{\rho} g_{\mu\nu} = \partial_\rho g_{\mu\nu} - 
\Gamma^\beta{}_{\rho \mu} g_{\beta \nu} -  \Gamma^\beta{}_{\rho \nu} g_{\mu 
\beta}  \,.
\end{equation}

When non-metricity is set to zero, the resulting geometry is Riemann-Cartan 
geometry. If torsion is set to zero, we recover torsion-free geometry. 
Similarly, setting curvature to zero leads to teleparallel geometry. Moreover, 
if both non-metricity and torsion are set to zero (resulting in the Levi-Civita 
connection), the geometry becomes Riemannian. On the other hand, if both 
non-metricity and curvature are set to zero (with the connection being the 
Weitzenb{\"{o}}ck one), the geometry is Weitzenb{\"{o}}ck. Lastly, when both 
curvature and torsion are set to zero (yielding the symmetric teleparallel 
connection), we obtain symmetric teleparallel geometry 
\cite{BeltranJimenez:2017tkd,Jarv:2018bgs}.

In Riemannian geometry, gravity is described using a Lagrangian based on the 
Ricci scalar, which is derived from contractions of the curvature tensor, 
leading to General Relativity. In contrast, in Weitzenb{\"{o}}ck geometry, 
gravity is described by a Lagrangian constructed from the torsion scalar, 
obtained through contractions of the torsion tensor, which corresponds to the 
Teleparallel Equivalent of General Relativity. Similarly, in symmetric 
teleparallel geometry, gravity is formulated through a Lagrangian involving 
contractions of the non-metricity tensor, specifically the non-metricity scalar
\begin{equation}
\label{NontyScalar}
Q=-\frac{1}{4}Q_{\alpha \beta \gamma}Q^{\alpha \beta 
\gamma}+\frac{1}{2}Q_{\alpha \beta \gamma}Q^{ \gamma \beta 
\alpha}+\frac{1}{4}Q_{\alpha}Q^{\alpha}-\frac{1}{2}Q_{\alpha}\tilde{Q}^{\alpha} 
\,,
\end{equation}
where
$Q_{\alpha}\equiv Q_{\alpha \ \mu}^{\ \: \mu} \, , $ and
$\tilde{Q}^{\alpha} 
\equiv Q_{\mu }^{\ \: \mu \alpha}  \, .
$
 
Inspired by the gravitational modifications in $f(R)$ and $f(T)$ theories, one 
can generalise the Lagrangian of the Symmetric Teleparallel Equivalent of 
General Relativity, $Q$, to an arbitrary function. This leads to the 
formulation 
of $f(Q)$ gravity, with the action given as
\cite{BeltranJimenez:2017tkd}, 
 \begin{equation}  
 \label{fQaction}
 S = -\frac{1}{2} \int {\mathrm{d}}^4 x \sqrt{-g}  f(Q) .
\end{equation}
Hence, Symmetric Teleparallel Equivalent of General Relativity is 
recovered for $f=Q/8\pi G$, where $8\pi G$ is the gravitational constant.

Variation of the total action $S+S_m$, where $S_m$  is the action 
corresponding to the matter sector,
leads to  the field equations, namely 
\cite{BeltranJimenez:2019tme, Dialektopoulos:2019mtr}: 
\begin{eqnarray}
&&  
\frac{2}{\sqrt{-g}} \nabla_{\alpha}\left\{\sqrt{-g} g_{\beta \nu} f_{Q} 
\left[- \frac{1}{2} L^{\alpha \mu \beta}+ \frac{1}{4} g^{\mu \beta} 
\left(Q^\alpha -  \tilde{Q}^\alpha \right) \right.\right.\nonumber\\
&&\left.\left. \ \ \ \ \ \ \ \ \ \ \ \ \ \ \ \ \ \  \ \ \ \ \ \ \ \ \ \ \ \, 
- \frac{1}{8} 
\left(g^{\alpha \mu} Q^\beta + g^{\alpha \beta} Q^\mu  
\right)\right]\right\}
 \nonumber \\
&& + f_{Q} \left[- \frac{1}{2} L^{\mu \alpha \beta}- \frac{1}{8} \left(g^{\mu 
\alpha} Q^\beta 
+ g^{\mu \beta} Q^\alpha  \right)
 \right. \nonumber\\
&&\left. \ \ \ \ \ \ \ + \frac{1}{4} g^{\alpha 
\beta} \left(Q^\mu -  \tilde{Q}^\mu \right)
\right] Q_{\nu \alpha 
\beta} +\frac{1}{2} \delta_{\nu}^{\mu} f=T_{\,\,\,\nu}^{\mu}\,,
\label{eoms}
\end{eqnarray}
with $f_{Q}=\partial f/\partial Q$, and $T_{\,\,\,\nu}^{\mu}$ the matter energy-momentum tensor, which in this kind of theories is conserved.

\subsection{\texorpdfstring{$f(Q)$} {f(Q)} cosmology}

In the previous subsection, we introduced $f(Q)$ gravity. To explore its 
application within a cosmological framework, we consider a flat Friedmann-Lemaître-Robertson-Walker (FLRW) metric given by:
\begin{equation}
\label{FRWmetric}
ds^{2}=-dt^{2}+a^{2}(t)\delta_{ij}dx^{i}dx^{j}\,,
\end{equation}
where $a(t)$ is the scale factor. It is worth mentioning that the lapse 
function 
can still be set to 1, as the non-metricity scalar $Q$ maintains a residual 
time-reparameterization invariance, even though the diffeomorphisms have been 
used to fix the coincident gauge \cite{BeltranJimenez:2018vdo,BeltranJimenez:2019tme}.
Furthermore, for the matter content, we assume it to be described by a perfect 
fluid with energy density $\rho_m$ and pressure $p_m$.
 
As it has been discussed in the literature, requiring  the connection to be symmetric, flat and to satisfy the FLRW symmetries, leads to three different  
connection classes, whose non-zero  
components are \cite{Hohmann:2020zre,Hohmann:2021ast,DAmbrosio:2021pnd,Heisenberg:2022mbo,
Paliathanasis:2023nkb,Shi:2023kvu,Basilakos:2025olm,Ayuso:2025vkc} 
\begin{align}
\Gamma _{~tt}^{t}& =K_{1}\,,~~\Gamma _{~rr}^{t}=K_{2}\,,~~\Gamma _{~\theta
\theta }^{t}=K_{2}r^{2}\,,  \notag  \label{Generalconnection} \\
\Gamma _{~tr}^{r}& =\Gamma _{~rt}^{r}=\Gamma _{~t\theta }^{\theta }=\Gamma
_{~\theta t}^{\theta }=\Gamma _{~t\phi }^{\phi }=\Gamma _{~\phi t}^{\phi
}=K_{3}\,,  \notag \\
\Gamma _{~r\theta }^{\theta }& =\Gamma _{~\theta r}^{\theta }=\Gamma
_{~r\phi }^{\phi }=\Gamma _{~\phi r}^{\phi }=\frac{1}{r}\,,~~\Gamma
_{~\theta \theta }^{r}=-r\,,  \notag \\
~~\Gamma _{~\phi \phi }^{r}& =-r\sin ^{2}\theta \,,~~\Gamma _{~\phi \phi
}^{t}=K_{2}r^{2}\sin ^{2}\theta \,,  \notag \\
\Gamma _{~\phi \theta }^{\phi }& =\Gamma _{~\theta \phi }^{\phi }=\cot
\theta \,,~~\Gamma _{~\phi \phi }^{\theta }=-\sin \theta \cos \theta \,,
\end{align}%
with  $K_{1}(t),K_{2}(t),K_{3}(t)$   functions of time given by
\begin{eqnarray}
&& \!\!\!\!\!\!\!\!\!\!\!\!\!\! \!\!\! \text{Connection 
I}\!:K_{1}=\gamma(t),K_{2}=0,K_{3}=0,
\\
&& \!\!\!\!\!\!\!\!\!\!\!\!\!\! \!\!\!\text{Connection II}\!:  
K_{1}=\frac{\dot{\gamma}(t)}{\gamma(t)}%
+\gamma(t),K_{2}=0,K_{3}=\gamma(t),\label{Connection2}
\\
 && \!\!\!\!\!\!\!\!\!\!\!\!\!\! \!\!\! \text{Connection III}\!:  
K_{1}=-\frac{\dot{\gamma}(t)}{\gamma(t)}%
,K_{2}=\gamma(t),K_{3}=0,
\end{eqnarray}
where $\gamma(t)$ a function of time.  Connection I
is the one that has been studied in detail in   cosmological analyses, since the other connections have the additional complication of the $\gamma(t)$ function. Hence, in this work we also focus on this connection choice, mainly since it allows for a more convenient perturbative analysis. 

It is worth stressing that the evolution and growth history in $f(Q)$ cosmology 
are not uniquely determined by the action and the FLRW metric alone: several 
inequivalent connections are consistent with the required symmetries, and their 
differences are encapsulated in the function $\gamma(t)$. This situation is not 
unfamiliar in theories with non-standard connections (for instance, in Palatini 
$f(R)$ gravity the independent connection couples algebraically to matter, 
leading to dynamics not controlled solely by the gravitational action).  

In the present work we restrict ourselves to the connection that reproduces the 
most widely studied case in the literature, for which the background field 
equations reduce to the familiar Friedmann-type equations in the coincident 
gauge~\cite{BeltranJimenez:2019tme}. In this case $\gamma(t)$ does not enter the 
background equations and no additional freedom affects the cosmological 
dynamics, which allows for a clearer comparison with previous studies.  

For more general connections, however, $\gamma(t)$ does enter the field 
equations, effectively providing the theory with   additional degree of 
freedom. Recent works have shown that this connection degree of freedom can 
combine with the scalar arising from the non-linear $f(Q)$ form to yield two 
scalar degrees of freedom, corresponding to the standard ``quintom'' 
cosmology~\cite{Basilakos:2025olm}. Furthermore, data-driven reconstructions of 
$\gamma(t)$ in $f(Q)$ cosmology have already been explored~\cite{Yang:2024tkw}.  

Importantly, the recovery of the GR limit is unaffected by the connection: in 
the limit $f(Q)\to Q-\Lambda$, one always obtains the Einstein field equations 
with a cosmological constant, independently of the connection considered 
\cite{Heisenberg:2023lru}. A recent and detailed analysis of the role of the affine connection in $f(Q)$ theories can be found in \cite{Ayuso:2025vkc}.

\subsubsection{Background evolution}

Under these   considerations, the general field equations (\ref{eoms})
lead to the two Friedmann equations, namely
\begin{eqnarray}
&&6f_QH^2-\frac{1}{2}f = \rho_m + \rho_r, \label{eqFrid}\\
&&\big(12H^2f_{QQ}+f_Q\big)\dot{H} = -\frac{1}{2} \big[(\rho_m+p_m) + (\rho_r+p_r)\big], \:\:\:\:\:\:\:\label{Fr2}
\end{eqnarray}
where $\rho_m$ and $\rho_r$ denote the energy densities of matter and radiation, respectively, and $p_m$ and $p_r$ their corresponding pressures. Moreover, in the context of FLRW geometry, the non-metricity scalar $Q$ from 
(\ref{NontyScalar}) simplifies to
\begin{eqnarray}
\label{QFRW}
Q=6H^2,
\end{eqnarray}
where $H\equiv\dot{a}/a$  is  the Hubble function.
As we observe, in $f(Q)$ 
cosmology, an effective dark energy component naturally 
emerges from the geometric modifications introduced by non-metricity. Finally, 
the system of equations is completed by incorporating the conservation 
equations for the matter and radiation fluids:
\begin{eqnarray}
\dot{\rho}_m + 3H(\rho_m + p_m) = 0, \quad
\dot{\rho}_r + 3H(\rho_r + p_r) = 0. \:\:\:\:\:\:\:\:\:\: \label{conservation}
\end{eqnarray}  
with $p_m = 0$ for pressureless matter and $p_r = \rho_r/3$ for radiation. The corresponding energy densities evolve as
\begin{equation}
\rho_m = \rho_{m0} a^{-3}, \qquad
\rho_r = \rho_{r0} a^{-4}, \label{conservationevolution}
\end{equation}
where $\rho_{m0}$ and $\rho_{r0}$ are the present-day energy densities for matter and radiation, respectively.

Since, in the case of FLRW geometry, the non-metricity scalar (\ref{QFRW}) 
happens 
to coincide with the torsion scalar in $f(T)$ gravity (where $T=6H^2$ in the 
mostly-plus metric convention), it follows that $f(Q)$ gravity and $f(T)$ 
gravity are equivalent at the background level.
However, because non-metricity and torsion generally possess distinct geometric 
structures, the perturbative behaviour of the two theories will differ, even 
within the highly symmetric framework of FLRW geometry, as we demonstrate in the 
next subsection.

\subsubsection{Scalar perturbations}
  
In this subsection, we discuss the scalar perturbations and derive the 
corresponding equation governing the growth of matter overdensities, following 
\cite{BeltranJimenez:2019tme}. By working within the coincident gauge and using 
conformal time $\tau$, the perturbed metric takes the form:
\begin{eqnarray}
\frac{d s^2}{a^2(\tau)} 
=  -\left( 1+ 2\phi\right)d \tau^2 + 2\left( B_{,i} \right) d\tau d x^i 
\nonumber \\
 +\Big[ \left( 1-2\psi\right)\delta_{ij}+ 2E_{,ij}   \Big]d 
x^id x^j\,, \label{conformal}
\end{eqnarray}
with $\phi$, $B$, $\varphi$ and $E$   the scalar perturbations, and where $\psi=\varphi+\frac{1}{3}\delta^{ij}E_{,ij}$ \cite{BeltranJimenez:2019tme}. 
Additionally, we perturb the 
\par\noindent\makebox[\columnwidth]{\rule{8.75cm}{0.2pt}}
\newpage
\par\noindent\makebox[\columnwidth]{\rule{8.75cm}{0.2pt}}
perfect-fluid matter energy-momentum tensor as
\ba
T^0{}_0 &= & -\rho_m\left(1+\delta\right)\,, \\
T^0{}_i &= & -\left(\rho_m+p_m\right)  \partial_iv    \,, \\
T^i{}_0 &= & \left(\rho_m+p_m\right)  \partial^i(v-B)    \,, \\
T^i{}_j & = & \left( p+\delta p_m\right)\delta^i_j  \,,
\ea
where for simplicity we neglect the anisotropic stress.
In the above expressions, $\rho_m$ and $p_m$ are the background energy density and 
pressure, and $\delta\equiv\delta\rho_m/\rho_m$ is the matter overdensity.

By substituting the aforementioned perturbations into the general field 
equations (\ref{eoms}), we arrive at \cite{BeltranJimenez:2019tme}

\ba
-a^2\delta\rho_m &=& 6\left( f_{Q} + 12a^{-2}\mH^2 f_{QQ}\right)\mH\left( \mH\phi 
+ \varphi'  \right) + 2f_{Q} k^2\psi  \nonumber\\
&&- 2\left[ f_{Q} + 3a^{-2}f_{QQ}\left(\mH'+\mH^2\right)\right]\mH k^2B\,,
\label{eq00}
\ea
\ba
&&\frac{1}{2}a^2\left(\rho_m+p_m\right) v = \left[ f_{Q} + 3a^{-2}f_{QQ}\left( 
\mH'+\mH^2\right)\right]\mH \phi   \nonumber\\
&&+ 6a^{-2}f_{QQ}\mH^2\varphi' - 9a^{-2}f_{QQ}\left(\mH'-\mH^2\right)\mH\varphi 
 \nonumber\\
&&+  f_{Q}\psi'-a^{-2}f_{QQ}\mH^2 k^2B\,, \label{eq0i}
\ea 

\onecolumngrid

\ba
\frac{1}{2}a^2\delta p_m & = & \left( f_{Q} + 12a^{-2}f_{QQ} \mH^2\right) \left( 
\mH\phi' + \varphi''\right) + 
\left[ f_{Q}\left( \mH'+2\mH^2-\frac{1}{3}k^2\right) + 
12a^{-2}f_{QQ}\mH^2\left( 4\mH'-\mH^2\right) + 12 a^{-2}\frac{d f_{QQ}}{d 
\tau}\mH^3\right]\phi \nn \\          
& + & 2\left[ f_{Q} + 6a^{-2}f_{QQ}\left(3\mH'-\mH^2\right) + 6a^{-2}\frac{d 
f_{QQ}}{d \tau}\mH\right]\mH\varphi'  + \frac{1}{3} f_{Q}k^2\psi \nn \\
& - &  \frac{1}{3}\left( f_{Q} + 6a^{-2}f_{QQ}\mH^2\right) k^2 B'
- \frac{1}{3}\left[ 2f_{Q}+3a^{-2}f_{QQ}\left( 5\mH-\mH^2\right) + 
6a^{-2}\frac{d f_{QQ}}{d \tau}\mH\right] \mH k^2B\,, \label{eqii}
\ea
\twocolumngrid
\hspace{-1em}with primes denoting differentiation with respect to the conformal time $\tau$, 
and where
$\mH\equiv a'/a=aH$ is the conformal Hubble function, $k$  
is the wavenumber of Fourier modes, and $f_{QQ}=\partial^2 f/\partial Q^2$. 
We can additionally define the matter equation-of-state parameter $w = p_m / 
\rho_m$, along with the adiabatic sound speed squared $c_s^2 = p_m' / \rho_m'$. Using these 
definitions, the continuity and Euler equations take the following form:
\ba
&&
\!\!\!\!\!\!\!\!\!\!\!\!\!\!
\delta'  =   \left( 1\!+\!w\right)\left(3\varphi'-k^2v-k^2B\right) + 3\mH\left(\! 
w\rho_m\!-\!\frac{\delta p_m}{\rho_m}\!\right), \label{eqcc0} \\
&&
\!\!\!\!\!\!\!\!\!\!\!\!\!\!
v'   =   -\mH\left( 1-c_s^2\right) v+ \frac{\delta p_m}{\rho_m + p_m}  + \phi\,. 
\qquad \label{eqcci}
\ea 
\par\noindent\makebox[\columnwidth]{\rule{8.75cm}{0.2pt}}
\newpage
\par\noindent\makebox[\columnwidth]{\rule{8.75cm}{0.2pt}}

In addition to the above, one must also account for the contributions stemming 
from the symmetric teleparallel connection structure and its perturbations, as 
given in \cite{BeltranJimenez:2019tme}:
\ba
&-& f_{QQ} \mH \left[ 2\mH \varphi' + \left( \mH' + \mH^2\right)\phi + \left( 
\mH'-\mH^2\right)\left( \psi - B'\right) \right] \nn \\ 
&-& \Big[  f_{QQ} \left( \mH'{}^2 +\mH \mH'' - 3\mH^2 \mH' -\frac{1}{3}\mH^2 
k^2\right) \nn \\ &+& 
 \frac{d f_{QQ}}{d \tau} (\mH'-\mH^2)\mH\Big]B = 0\,, \label{eqc0}
\ea
and
\onecolumngrid
\ba
&- &f_{QQ} \left( \mH'-3\mH^2\right)\mH\phi'  -  
\left[ f_{QQ}\left( \mH''\mH + \mH'{}^2 - 9\mH' \mH^2\right) - \frac{d 
f_{QQ}}{d \tau}\left(\mH'-3\mH^2\right)\mH\right] \phi \nn \\
+\quad 2f_{QQ}\mH^2\varphi'' & + & 
 \left[ f_{QQ} \left( \mH'+3\mH^2\right)  +  2\frac{d f_{QQ}}{d \tau}\mH 
\right]\mH\varphi'  -3   
\left[ f_{QQ} \left(\mH'{}^2+\mH''\mH-3\mH'\mH^2\right)  + \frac{d f_{QQ}}{d 
\tau}\left(\mH'-\mH^2\right)\mH\right]\varphi \nn \\
- \quad \frac{1}{3}f_{QQ} \mH^2 k^2B' & + & \frac{1}{3}\left[ 
f_{QQ}\left(\mH'-3\mH^2\right) - \frac{d f_{QQ}}{d \tau} \mH\right]\mH k^2 B 
=0\,. \label{eqci}
\ea
\newpage
\twocolumngrid

In summary, as observed, compared to standard curvature-based modified 
gravity, 
there are two extra metric perturbation variables accompanied by two additional 
equations.

Next, we focus on sub-horizon scales, where $k \gg \mH$, as this is the regime 
where matter clustering occurs. As is customary, we consider dust-like matter 
with full clustering properties, characterised by ($w=0$ and $c_s^2=0$). Under 
these conditions, equation (\ref{eq00}) reduces to the Poisson equation:
\be \label{newton}
\psi = -\frac{4\pi G\rho_m \delta}{k^2f_{Q}}\,.
\ee
Moreover, note that in this case   (\ref{eqii})  gives 
\be
\phi = \psi\,,
\label{phipsi}
\ee
as expected. Hence, equation
(\ref{eqc0}) 
simplifies to 
 \be \label{betasol}
B = \frac{6}{k^2}\left( \varphi' + \frac{\mH'}{\mH}\phi\right)\,.
\ee
On the other hand, the continuity equation (\ref{eqcc0}) becomes
$
\delta'' = -k^2\left( v' +B'\right) + 3\varphi''\,,
$
which,   using (\ref{eqc0}) and (\ref{eqci}) finally leads to
\be
\delta'' + \mH\delta' + k^2\phi = 3\left( \varphi'' + \mH\varphi'\right) 
-k^2\left( B'+\mH'B\right)\,.
\label{helpequa}
\ee

Therefore, to maintain consistency with (\ref{newton}) and (\ref{betasol}), as 
well as with the small-scale approximation (quasi-static limit), we can 
disregard the right-hand side of equation (\ref{helpequa}), which simplifies to:
\be
\delta'' + \mH\delta' =  \frac{4\pi G\rho_m}{f_{Q}} \delta\,,
\label{finalequation}
\ee
where we have also used (\ref{phipsi}).

 Equation (\ref{finalequation}) governs the evolution of matter overdensity on 
sub-horizon scales for standard dust matter. The key distinction compared to 
General Relativity lies in the presence of an effective Newton's constant, 
given 
by:
\be
G_{\textrm{eff}} \equiv  \frac{  G}{f_{Q}} \,.
\label{Geff}
\ee
This effective Newton's constant is a common feature in the corresponding 
equations of many modified gravity theories, serving as a measure of how 
gravitational modifications influence the growth of matter clustering \cite{Basilakos:2012uu}.

\subsection{Specific \texorpdfstring{$f(Q)$}{f(Q)} models}
\label{fQmodels}

We now introduce and discuss the specific $f(Q)$ models considered in this work. Before presenting them, let us recall that in the linear case $f(Q)=Q/8\pi G$, the theory reduces to the Symmetric Teleparallel Equivalent of General Relativity (STGR). Likewise, if a constant term is included, $f(Q)=Q/8\pi G - 2\Lambda$, the framework exactly reproduces the standard $\Lambda$CDM cosmology. Furthermore, as shown in \cite{BeltranJimenez:2019tme}, one can add to the Lagrangian a $\sqrt{Q}$ term without affecting the background evolution, in analogy with $f(T)$ gravity. 

In this context, we focus on three non-linear $f(Q)$ models that give rise to non-trivial cosmological phenomenology, while remaining simple enough for analytical treatment. Their explicit forms are:
\begin{align}
f_1(Q) &= Q\,e^{\lambda_1 Q_0/Q}, \label{fqone} \\
f_2(Q) &= Q+Q_0 e^{-\lambda_2 Q_0/Q}, \label{fqtwo} \\
f_3(Q) &= Q+\lambda_3 Q_0\!\left[1-e^{-Q_0/Q}\right], \label{BGeqsfQ}
\end{align}
where $\lambda_i$ is the dimensionless parameter associated with each model, and $Q_0 \equiv 6H_0^2$ with $H_0$ the present Hubble constant. 

Throughout the paper, we consider a background composed of one pressureless matter component ($w_m = 0$) and one radiation component ($w_r = 1/3$). Hence, the Friedmann-like equations presented below include both matter and radiation contributions, $\Omega_{m0} a^{-3}$ and $\Omega_{r0} a^{-4}$, respectively. 

It is important to stress that the above functional forms are not intended to 
span the full space of possible $f(Q)$ models, but rather to serve as simple, 
physically motivated representatives. They satisfy two basic requirements:  
(i)~all three reduce to the GR limit at early-time ($E\gg 1$, $a\ll 1$), with 
$f(Q)\to Q$ and $f_Q\to 1$, thereby ensuring consistency with the well-tested 
radiation- and matter-dominated eras;  
(ii)~they constitute minimal one-parameter deformations of GR built from 
exponential functions of $Q_0/Q$, which act as resummations of inverse-$Q$ 
corrections and naturally generate late-time acceleration without introducing a 
cosmological constant.  
Additionally, these ansätze illustrate qualitatively distinct possibilities at 
late-time, covering phantom-like ($w_{\mathrm{DE}}<-1$) versus 
quintessence-like ($w_{\mathrm{DE}}>-1$) dynamics, as well as enhanced 
($G_{\mathrm{eff}}>G$) versus suppressed ($G_{\mathrm{eff}}<G$) effective 
Newton’s constant, as we will see in the following paragraphs.  

Finally, we note that $f(Q)$ gravity differs from $f(R)$ gravity in that the 
field equations remain second order rather than fourth order. This simpler 
structure allows for greater flexibility in the choice of functional forms. 
Nevertheless, certain physical conditions must still be respected, such as the 
positivity of $f_Q$ (ensuring $G_{\mathrm{eff}}>0$) and the recovery of GR in 
the appropriate limit, all of which are satisfied by the three models 
considered here.


\subsubsection*{Model 1: Exponential $f_1(Q)$}

For $f_1(Q)$ the normalised Hubble function $E^2 \equiv H^2/H_0^2$ satisfies

\begin{equation}\label{friedmodeli}
(E^2 - 2\lambda_1) e^{\lambda_1/E^2} = \Omega_{m0} a^{-3} + \Omega_{r0} a^{-4},
\end{equation}
with $\lambda_1$ given by

\begin{equation}\label{lambdamodeli}
\lambda_1 = \frac{1}{2} + \mathcal{W}_0\!\left(-\frac{\Omega_{m0}+\Omega_{r0}}{2\sqrt{e}}\right),
\end{equation}
where $\mathcal{W}_0$ is the principal branch of the Lambert function. The corresponding effective Newton’s constant reads

\begin{equation}\label{geff1}
   G_{\textrm{eff}}=\frac{G}{e^{\lambda_1/E^2}(1-\lambda_1/E^2)}.
\end{equation}

\subsubsection*{Model 2: Exponential $f_2(Q)$}

For $f_2(Q)$ the Hubble function satisfies

\begin{equation}\label{friedmodelii}
E^2 + e^{-\lambda_2/E^2}\!\left(2\lambda_2/E^2-1\right)
= \Omega_{m0} a^{-3} + \Omega_{r0} a^{-4},
\end{equation}
with

\begin{equation}\label{lambdamodelii}
\lambda_2 = \frac{1}{2}-\mathcal{W}_0\!\left[ -\frac{\sqrt{e}}{2}\left(\Omega_{m0}+\Omega_{r0}-1\right)\right],
\end{equation}
and

\begin{equation}\label{geff2}
G_{\textrm{eff}}=\frac{G}{1+\lambda_2 E^{-4}e^{-\lambda_2/E^{2}}}.
\end{equation}

\subsubsection*{Model 3: Novel exponential $f_3(Q)$}

Finally, we propose a new form,

\begin{equation}
f_3(Q) = Q+\lambda_3 Q_0\left[1-\exp(-Q_0/Q)\right],
\end{equation}
leading to

\begin{equation}\label{friedmodeliii}
E^2 + \lambda_3 e^{-1/E^2}\!\left(1-2/E^2\right) - \lambda_3
= \Omega_{m0}a^{-3} + \Omega_{r0} a^{-4},
\end{equation}
with

\begin{equation}\label{lambdamodeliii}
\lambda_3 = \frac{e}{1+e}\left(1-\Omega_{m0}-\Omega_{r0}\right),
\end{equation}
and

\begin{equation}\label{geff3}
G_{\textrm{eff}}=\frac{G}{1-\lambda_3 E^{-4}e^{-1/E^2}}.
\end{equation}

\medskip
To further interpret the phenomenology of these models, it is useful to introduce effective quantities that characterise both the background expansion and the late-time acceleration. The first is the total effective equation of state, 

\begin{equation}
w_{\mathrm{tot}}\equiv \frac{p_{\mathrm{tot}}}{\rho_{\mathrm{tot}}},
\end{equation}
with

\begin{equation}
p_{\mathrm{tot}}=-\frac{2\dot{H}+3H^2}{8\pi G}, \qquad \rho_{\mathrm{tot}}=\frac{3H^2}{8\pi G}.
\end{equation}
This leads to

\begin{equation}
w_{\mathrm{tot}} = -1 - \frac{2\dot{H}}{3H^2},
\end{equation}
which describes the global expansion history. The second is the effective dark energy equation of state,

\begin{equation}
w_{\mathrm{DE}} \equiv \frac{p_{\mathrm{DE}}}{\rho_{\mathrm{DE}}},
\end{equation}
where we define $p_{\mathrm{DE}}\equiv p_{\mathrm{tot}}-p_m-p_r$ and $\rho_{\mathrm{DE}}\equiv \rho_{\mathrm{tot}}-\rho_m-\rho_r$. Here $p_m=0$ and $p_r=\rho_r/3$, while

\begin{equation}
\rho_m=\frac{3H_0^2}{8\pi G}\Omega_{m0}(1+z)^3, \qquad 
\rho_r=\frac{3H_0^2}{8\pi G}\Omega_{r0}(1+z)^4,
\end{equation}
where $z$ denotes the redshift, related to the scale factor by $1+z=a^{-1}$. Hence,

\begin{equation}
w_{\mathrm{DE}}=-\frac{2\dot{H}+3H^2+H_0^2\Omega_{r0}(1+z)^4}{3\left[H^2-H_0^2\left(\Omega_{m0}(1+z)^3+\Omega_{r0}(1+z)^4\right)\right]}.
\end{equation}

\begin{figure*}[t]
    \centering
    \begin{tabular}{cc}
        \includegraphics[width=0.5\textwidth]{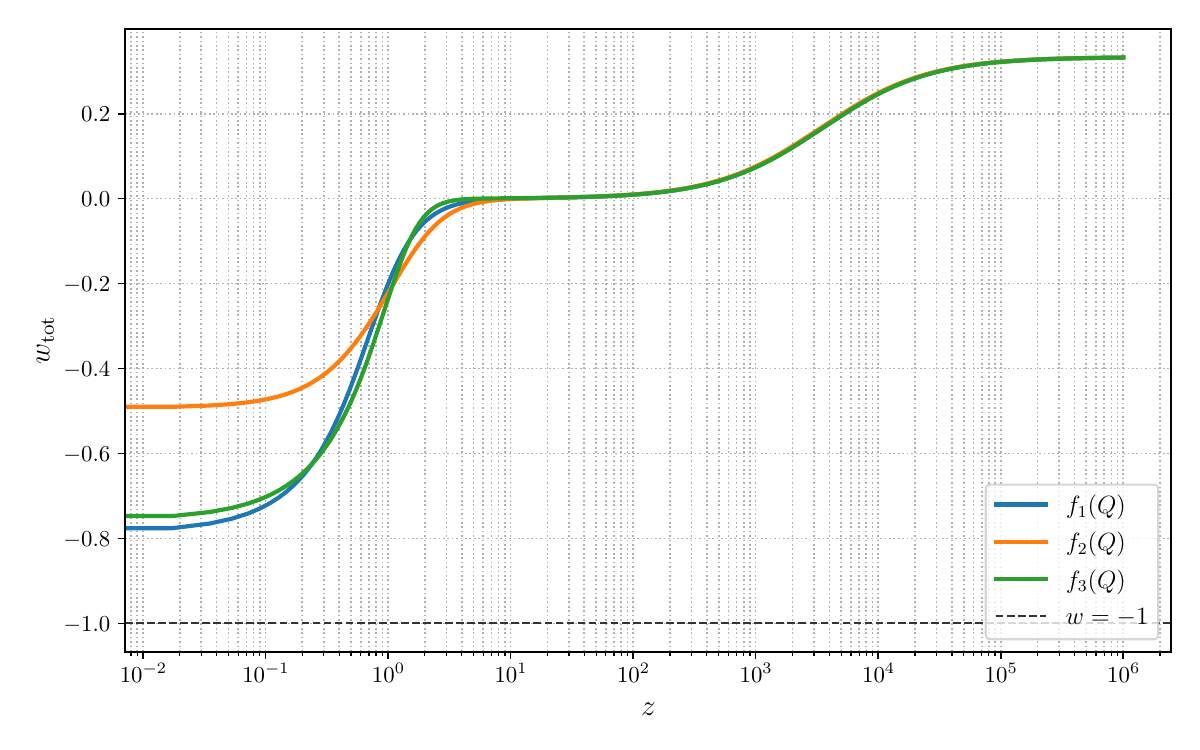} &
        \includegraphics[width=0.5\textwidth]{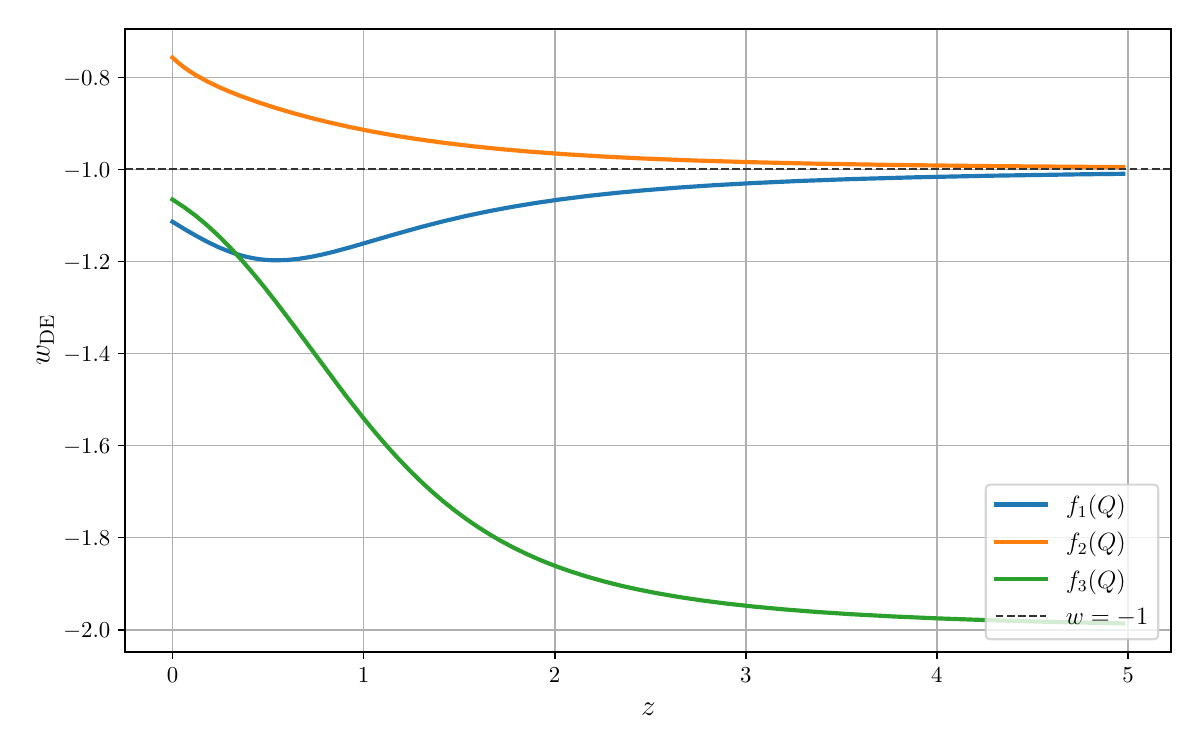} \\
        \multicolumn{2}{c}{\includegraphics[width=0.6\textwidth]{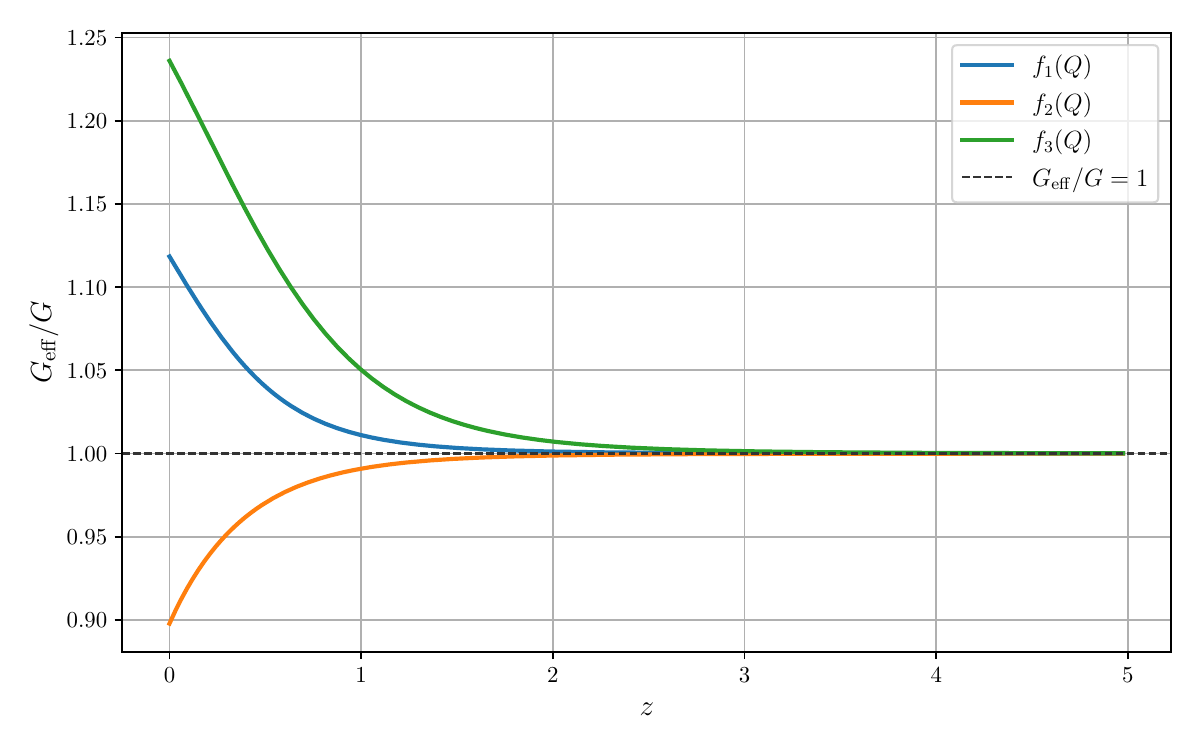}} \\
    \end{tabular}
    \caption{\justifying{\it{
    Cosmological background and effective properties of the $f(Q)$ models compared to $\Lambda$CDM.
    \textbf{Top-left:} Total effective equation of state $w_{\mathrm{tot}}(z)$, 
    showing the transition from radiation to matter domination and the late-time 
    accelerated regime. 
    \textbf{Top-right:} Effective dark energy equation of state $w_{\mathrm{DE}}(z)$ 
    compared to $\Lambda$CDM ($w=-1$). 
    \textbf{Bottom:} Effective Newton’s constant $G_{\mathrm{eff}}/G$ as a function of redshift. 
    Models~1 and~3 exhibit $G_{\mathrm{eff}}>G$, whereas Model~2 shows $G_{\mathrm{eff}}<G$, 
    with distinct implications for structure formation. The horizontal line $G_{\mathrm{eff}}/G=1$ marks the GR limit. All curves are obtained using the best-fit cosmological parameters corresponding to Combination~V, which will be defined in the results section.}}}
    \label{fig:theory_motivation}
\end{figure*}

As discussed above, all three $f(Q)$ models reduce to the GR limit at early-time 
($E \gg 1$, $a \ll 1$), reproducing the standard sequence of radiation and matter 
domination before entering the accelerated expansion phase (top-left panel of 
Fig.~\ref{fig:theory_motivation}). The differences among the models appear only 
at late-time, where the accelerated regime is reached. In this epoch, a variety 
of behaviours emerges: Models~1 and~3 exhibit phantom-like dynamics, as illustrated 
in the evolution of the effective dark energy equation of state $w_{\mathrm{DE}}(z)$ 
(top-right panel of Fig.~\ref{fig:theory_motivation}), while Model~2 follows a 
quintessence-like trajectory. Similarly, both cases $G_{\mathrm{eff}}/G>1$ and 
$G_{\mathrm{eff}}/G<1$ are realised within the considered models, with Models~1 
and~3 corresponding to the former and Model~2 to the latter, as displayed in the 
bottom panel of Fig.~\ref{fig:theory_motivation}. We remark that achieving a 
simultaneous phantom-like behaviour together with $G_{\mathrm{eff}}/G<1$ is highly 
non-trivial in $f(Q)$ cosmology, as in many other modified gravity frameworks. 
This issue will be revisited in the discussion of our results.

\section{Observational constraints}
\label{Observations}

\subsection{Data and  Methodology}

We utilise Cobaya’s Monte Carlo Markov Chain (MCMC) sampler \cite{Torrado:2020dgo} to generate the posterior distribution of the full cosmological parameter space. The MCMC runs are performed in single-chain mode, and convergence is assessed using the default Gelman–Rubin $R-1$ statistics implemented in Cobaya, following the prescription of \cite{Lewis:2013hha}. Our analysis is fully Bayesian, based on the likelihood function $\mathcal{L}_{\textrm{tot}} \sim \textrm{exp}(-\chi^2_{\textrm{tot}}/2)$, where the total chi-squared, $\chi^2_{\textrm{tot}}$, is computed as the sum of contributions from various cosmological probes: $\chi^2_{\textrm{CC}}, \chi^2_{\textrm{SN}}, \chi^2_{\textrm{GRB}}, \chi^2_{\textrm{BAO}}, \chi^2_{\textrm{CMB}}$, each of which will be defined and discussed in detail in the following paragraphs.

\Needspace{8\baselineskip}

The parameter space explored in this work is described by the vector of free parameters

\begin{equation}\label{thetadef}
\bm{\theta} =
\begin{cases}
\{H_0,\,\Omega_{m0}\}, & \text{for Combination I},\\[4pt]
\{H_0,\,\Omega_{m0},\,\Omega_{b0}\},
  & \text{\parbox[t]{.55\linewidth}{for the rest of combinations,}}
\end{cases}
\end{equation}
where $H_0$ is the present-day Hubble constant, $\Omega_{m0}$ is the present-day matter density parameter, 
and $\Omega_{b0}$ is the present-day baryon density parameter \footnote{We do not treat $\Omega_{r0}$ as a free parameter. Instead, it is derived from $\Omega_{m0}$ and $H_0$ through the relations $h = H_0 / 100$, $z_{\mathrm{eq}} = 2.5 \times 10^4\, \Omega_{m0} h^2 (T_{\mathrm{CMB}}/2.7)^{-4}$, and $\Omega_{r0} = \Omega_{m0}/(1+z_{\mathrm{eq}})$ \cite{Eisenstein:1997ik}. We set $T_{\mathrm{CMB}}=2.7255\,\mathrm{K}$ \cite{Fixsen:2009ug}.}.
The precise definitions of the different combinations will be introduced when presenting the results. Finally, we adopt flat priors on all parameters, summarised in Tab.~\ref{tab:priors}.

\subsubsection{Cosmic Chronometers}\label{ccdata}

We utilise measurements of the Hubble parameter $H(z)$ obtained from cosmic chronometers (CC), which provide a model-independent method for understanding the expansion history of the universe. CC are passively evolving massive galaxies whose ages can be used to trace the expansion history of the universe \cite{Jimenez:2001gg}. This method is particularly valuable as it allows direct measurements of $H(z)$ without assuming a specific cosmological model.

We make use of a dataset that spans a redshift range of $z\sim0.07$ to $z\sim2.0$, comprising $33$ measurements of $H(z)$ \cite{Moresco:2024wmr}. These data points are provided in the form $(z_i,H_i,\sigma_{H_i})$, where $z_i$ represents the redshift, $H_i$ is the Hubble parameter at $z_i$, and $\sigma_{H_i}$ denotes the observational uncertainty. These measurements have been used extensively in the literature to constrain cosmological parameters and, in this study, we incorporate them into our analysis to derive constraints on the model parameters of interest.

CC data are incorporated into the analysis via a chi-squared statistics, defined as
\begin{equation}\label{chicc}
\chi^2_{\textrm{CC}}=\sum_{i=1}^{33} \frac{[H_{\textrm{obs}}(z_i)-H_{\textrm{th}}(z_i,\bm{\theta})]^2}{\sigma^2_{H_i}},
\end{equation}
where $H_{\textrm{th}}(z_i,\bm{\theta})$ represents the theoretical value of the Hubble parameter at redshift $z_i$, based on the cosmological parameters $\bm{\theta}$, and $H_{\textrm{obs}}(z_i)$ is the observed value.

\begin{table}[t]
    \centering
    \begin{tabular}{@{\hskip 15pt}c@{\hskip 15pt}c@{\hskip 15pt}c@{\hskip 15pt}}
        \hline
        \hline
        \addlinespace[2pt]
        Param. & Min. & Max. \\
        \addlinespace[2pt]
        \hline
        \hline
        \addlinespace[2pt]
        $H_0$  & $40$ & $100$ \\
        \addlinespace[2pt]
        \hline
        \addlinespace[2pt]
        $\Omega_{m0}$  & $0.1$ & $0.9$ \\
        \addlinespace[2pt]
        \hline
        \addlinespace[2pt]
        $\Omega_{b0}$  & $0.01$ & $0.09$ \\
        \addlinespace[2pt]
        \hline
        \hline
    \end{tabular}
    \caption{\justifying{Flat priors adopted for the cosmological parameters in the MCMC analysis. The ranges shown define the minimum and maximum values allowed for each parameter. For Combination I, only $H_0$ and $\Omega_{m0}$ are considered, while $\Omega_{b0}$ is included in Combinations II, III, IV, and V.}}
    \label{tab:priors}
\end{table}

\subsubsection{Supernovae}\label{sndata}

We utilise the Pantheon sample of Type Ia supernovae (SN) \cite{Pan-STARRS1:2017jku} to constrain the cosmological model parameters. The Pantheon dataset is a compilation of 1048 spectroscopically confirmed Type Ia supernovae, spanning a redshift range from $z=0.01$ to $z=2.26$. These supernovae serve as standardised candles, providing precise measurements of the distance modulus, which is critical for understanding the expansion history of the universe.

The observed distance modulus $\mu_{\textrm{SN}}^{\textrm{obs}}$ for each supernova is expressed as
\begin{equation}\label{muobssn}
\mu_{\textrm{SN}}^{\textrm{obs}}(z)=m_B(z)-\mathcal{M},
\end{equation}
where $m_B$ is the apparent magnitude of the supernova, and $\mathcal{M}$ is the absolute magnitude, assumed to be constant for Type Ia supernovae. The observed distance modulus $\mu_{\textrm{SN}}^{\textrm{obs}}$ is compared to the theoretical prediction, $\mu_{\textrm{SN}}^{\textrm{th}}$, which depends on the cosmological model parameters through the luminosity distance:

\begin{equation}\label{muthsn}
\mu_{\textrm{SN}}^{\textrm{th}}(z_{\textrm{hel}},z_{\textrm{cmb}},\bm{\theta})=5\log_{10}[D_L(z_{\textrm{hel}},z_{\textrm{cmb}},\bm{\theta})]+\mu_0,
\end{equation}
where $\mu_0=5\log_{10}[cH_0^{-1}\textrm{Mpc}^{-1}]$, $z_{\textrm{cmb}}$ and $z_{\textrm{hel}}$ are, respectively, the CMB frame and heliocentric redshift \cite{SNLS:2011lii}, and $\bm{\theta}$ is the vector containing the cosmological parameters ($c$ is the speed of light). Note that we have used the Hubble-free luminosity distance $(H_0d_L/c)$, which is defined as
\begin{equation}\label{dlsn}
D_L(z_{\textrm{hel}},z_{\textrm{cmb}},\bm{\theta})=(1+z_{\textrm{hel}})\int_0^{z_{\textrm{cmb}}}dz'\frac{H_0}{H(z',\bm{\theta})}.
\end{equation}

In our analysis we include the full covariance matrix associated with the Pantheon sample, which accounts for statistical and systematic uncertainties. The covariance matrix ensures that correlations between supernovae at different redshift are appropriately incorporated, enhancing the robustness of our parameter constraints. The chi-squared statistics used to fit the supernova data is defined as
\begin{equation}\label{chisn}
\chi^2_{\textrm{SN}}=\bm{\Delta\mu_{\textrm{SN}}^T\cdot\C^{-1}_{\textrm{SN}}\cdot\Delta\mu_{\textrm{SN}}},
\end{equation}
where $\bm{\Delta\mu_{\textrm{SN}}}=\bm{\mu_{\textrm{SN}}^{\textrm{obs}}}-\bm{\mu_{\textrm{SN}}^{\textrm{th}}}$ represents the vector of residuals, i.e., the difference between the observed and theoretical distance moduli, and $\bm{\C^{-1}_{\textrm{SN}}}$ is the inverse covariance matrix. Note that $\mu_{\textrm{SN}}^{\textrm{obs}}-\mu_{\textrm{SN}}^{\textrm{th}}$ depends on the nuisance parameter $\widetilde{\mathcal{M}}$:
\begin{equation}\label{nuisancesn}
\mu_{\textrm{SN}}^{\textrm{obs}}-\mu_{\textrm{SN}}^{\textrm{th}}=m_B-5\log_{10}(D_L)-\widetilde{\mathcal{M}},
\end{equation}
where we have defined $\widetilde{\mathcal{M}}\equiv\mathcal{M}+\mu_0$. In fact, we can marginalise over it, obtaining \cite{SNLS:2011lii}
\begin{equation}\label{finalchisn}
\widetilde{\chi}^2_{\textrm{SN}}=A_{\textrm{SN}}+\ln{\frac{E_{\textrm{SN}}}{2\pi}}-\frac{B_{\textrm{SN}}^2}{E_{\textrm{SN}}},
\end{equation}
where $A_{\textrm{SN}}=\bm{\Delta\widetilde{\mu}_{\textrm{SN}}^T\cdot\C^{-1}_{\textrm{SN}}\cdot\Delta\widetilde{\mu}_{\textrm{SN}}}$, $B_{\textrm{SN}}=\bm{\Delta\widetilde{\mu}_{\textrm{SN}}^T\cdot\C^{-1}_{\textrm{SN}}\cdot1}$, \newline  $E_{\textrm{SN}}=\bm{1^T\cdot\C^{-1}_{\textrm{SN}}\cdot1}$, and $\bm{\Delta\widetilde{\mu}_{\textrm{SN}}}=\bm{\Delta\mu_{\textrm{SN}}}(\widetilde{\mathcal{M}}=0)$. Therefore, in our calculations we use the last expression for $\widetilde{\chi}^2_{\textrm{SN}}$.

\subsubsection{Gamma-Ray Bursts}\label{grbdata}

Gamma-ray bursts (GRB), some of the most energetic astrophysical phenomena, have gained prominence as potential cosmological probes that complement Type Ia supernovae. In particular, the GRB dataset known as the ``Mayflower'' sample \cite{Liu:2014vda}, which consists of 79 GRB spanning a redshift range of $1.44<z<8.1$, offers unique insights into the high-redshift universe.

For each GRB in the sample, the distance modulus $\mu_{\textrm{GRB}}^{\textrm{th}}$ is defined as

\begin{equation}\label{muthgrb}
\mu_{\textrm{GRB}}^{\textrm{th}}(z,\bm{\theta})=5\log_{10}[D_L(z,\bm{\theta})]+\mu_0,
\end{equation}
where $\mu_0=5\log_{10}[cH_0^{-1}\textrm{Mpc}^{-1}]$, and $\bm{\theta}$ is the vector containing the cosmological parameters. Analogously to the SN case, we have used the Hubble-free luminosity distance:

\begin{equation}\label{dlgrb}
D_L(z,\bm{\theta})=(1+z)\int_0^zdz'\frac{H_0}{H(z',\bm{\theta})}.
\end{equation}
These measurements extend to higher redshift than most other cosmological probes, offering valuable constraints on the expansion history of the universe.

The Mayflower sample’s GRB data are analysed using a chi-squared statistics, analogous to that for supernovae:
\begin{equation}\label{chigrb}
\chi^2_{\textrm{GRB}}=\bm{\Delta\mu_{\textrm{GRB}}^T\cdot\C^{-1}_{\textrm{GRB}}\cdot\Delta\mu_{\textrm{GRB}}},
\end{equation}
where $\bm{\Delta\mu_{\textrm{GRB}}}=\bm{\mu_{\textrm{GRB}}^{\textrm{obs}}}-\bm{\mu_{\textrm{GRB}}^{\textrm{th}}}$ represents the vector of residuals, and $\bm{\C^{-1}_{\textrm{GRB}}}$ is the inverse covariance matrix. In this case, we assume that the errors are uncorrelated, so the covariance matrix is given by the square of the errors: $\bm{\C_{\textrm{GRB}}}=\bm{\sigma^2_{\textrm{GRB}}}$. As happened in the SN case, $\mu_{\textrm{GRB}}^{\textrm{obs}}-\mu_{\textrm{GRB}}^{\textrm{th}}$ depends on a nuisance parameter, which is now $\mu_0$. Again, we can marginalise over it, obtaining \cite{SNLS:2011lii}
\begin{equation}\label{finalchigrb}
\widetilde{\chi}^2_{\textrm{GRB}}=A_{\textrm{GRB}}+\ln{\frac{E_{\textrm{GRB}}}{2\pi}}-\frac{B_{\textrm{GRB}}^2}{E_{\textrm{GRB}}},
\end{equation}
where $A_{\textrm{GRB}}=\bm{\Delta\widetilde{\mu}_{\textrm{GRB}}^T\cdot\C^{-1}_{\textrm{GRB}}\cdot\Delta\widetilde{\mu}_{\textrm{GRB}}}$, $B_{\textrm{GRB}}=\bm{\Delta\widetilde{\mu}_{\textrm{GRB}}^T\cdot\C^{-1}_{\textrm{GRB}}\cdot1}$,   $E_{\textrm{GRB}}=\bm{1^T\cdot\C^{-1}_{\textrm{GRB}}\cdot1}$, and $\bm{\Delta\widetilde{\mu}_{\textrm{GRB}}}=\bm{\Delta\mu_{\textrm{GRB}}}(\mu_0=0)$. We use the last expression for $\widetilde{\chi}^2_{\textrm{GRB}}$.

\subsubsection{Baryon Acoustic Oscillations}\label{baodata}

Baryon Acoustic Oscillations (BAO) appear as periodic fluctuations in the density of visible baryonic matter, serving as a cosmological standard ruler defined by the sound horizon radius at the drag epoch. The comoving sound horizon at redshift $z$ is given by 

\begin{equation}\label{rs_general}
    r_s(z,\bm{\theta}) = \int_{z}^\infty \frac{c_s(z')}{H(z',\bm{\theta})}\,dz',
\end{equation}
where $c_s(z)$ is the sound speed, expressed as

\begin{equation}\label{soundspeed}
    c_s(z) = \frac{c}{\sqrt{3\left[1+R_b(1+z)^{-1}\right]}},
\end{equation}
where $c$ is the speed of light and $R_b$ is given by \cite{Eisenstein:1997ik}

\begin{equation}
    R_b = 31500\Omega_{b0}h^2\left(\frac{T_{\textrm{CMB}}}{2.7}\right)^{-4}.
\end{equation}
Here, $\Omega_{b0}$ is the present-day baryon density parameter, $h=H_0/100$ and $T_{\textrm{CMB}}=2.7255$ K. For BAO analyses, the sound horizon is evaluated at the drag epoch, $z_d$. The drag redshift $z_d$ is computed using the fitting formula \cite{Eisenstein:1997ik}

\begin{equation}\label{zd}
    z_d = \frac{1291\left(\Omega_{m0}h^2\right)^{0.251}}{1+0.659\left(\Omega_{m0}h^2\right)^{0.828}}
    \left[1+b_1\left(\Omega_{b0} h^2\right)^{b_2}\right],
\end{equation}
with the coefficients

\begin{align}
    &b_1 = 0.313\left(\Omega_{m0}h^2\right)^{-0.419}\left[1+0.607\left(\Omega_{m0}h^2\right)^{0.6748}\right],\\
    &b_2 = 0.238\left(\Omega_{m0}h^2\right)^{0.223}.
\end{align}

We utilise the DESI BAO measurements \cite{DESI:2024mwx} obtained from various samples: The Bright Galaxy Sample (BGS, $0.1 < z < 0.4$), the Luminous Red Galaxy Sample (LRG, $0.4 < z < 0.6$ and $0.6 < z < 0.8$), the Emission Line Galaxy Sample (ELG, $1.1 < z < 1.6$), the combined LRG and ELG Sample (LRG+ELG, $0.8 < z < 1.1$), the Quasar Sample (QSO, $0.8 < z < 2.1$) and the Lyman-$\alpha$ Forest Sample (Ly$\alpha$, $1.77 < z < 4.16$). All these measurements are expressed in terms of $r_s(z_d,\bm{\theta})$ and a set of cosmological distance measures, namely the Hubble distance
\begin{equation}\label{hubbledist}
    D_H(z,\bm{\theta})=\frac{c}{H(z,\bm{\theta})},
\end{equation}
the comoving angular-diameter distance 
\begin{equation}\label{angulardist}
 D_M(z,\bm{\theta}) = \int_0^z \frac{c}{H(z',\bm{\theta})}dz',
\end{equation}
and the spherically-averaged distance
\begin{equation}\label{angledist}
D_V(z,\bm{\theta}) = \left[zD^2_M(z,\bm{\theta})D_H(z,\bm{\theta})\right]^{1/3}.
\end{equation}

We can now introduce the chi-squared statistics used to fit the BAO data as
\begin{equation}\label{chisqBAO} \chi^2_{\text{BAO}}= \bm{\Delta \textbf{X}^T_{\textrm{BAO}}\cdot \C^{-1}_{\textrm{BAO}}\cdot\Delta \textbf{X}_{\textrm{BAO}}}, 
\end{equation}
where $\bm{\Delta \textbf{X}_{\textrm{BAO}}} = \bm{\textbf{x}_{\textrm{BAO}}^{\textrm{obs}}}-\bm{\textbf{x}_{\textrm{BAO}}^{\textrm{th}}}$ denotes the residual vector between the observed and theoretical values of the BAO observables, and $\bm{\C^{-1}_{\textrm{BAO}}}$ is the inverse covariance matrix. In this work,  we include the full covariance matrix as provided by DESI, which is non-diagonal due to the inclusion of correlations among different redshift bins and observables.

\subsubsection{Cosmic Microwave Background}\label{cmbdata}

The Cosmic Microwave Background (CMB) provides a precise probe of the early Universe. Its constraining power for dark energy is efficiently captured by a set of distance priors derived from the angular scale of the sound horizon at photon decoupling. These priors provide a reliable summary of the information contained in the full CMB likelihood while remaining robust across a wide class of cosmological models. In this work we adopt the compressed CMB likelihood \cite{Zhai:2018vmm}, expressed in terms of the shift parameters $R$ and $\ell_a$, and the physical baryon density today $\Omega_{b0}h^2$. The shift parameters are given by

\begin{align}
    &R(z_\ast,\bm{\theta}) = \sqrt{\Omega_{m0}H_0^2}\,\frac{r(z_\ast,\bm{\theta})}{c},\\
    &\ell_a(z_\ast,\bm{\theta}) = \pi \frac{r(z_\ast,\bm{\theta})}{r_s(z_\ast,\bm{\theta})},
\end{align}
where $r(z_\ast,\bm{\theta})$ is the comoving distance evaluated at the redshift of photon decoupling $z_\ast$, $r(z_\ast,\bm{\theta})=D_M(z_\ast,\bm{\theta})$ (see Eq. \eqref{angulardist}), and $r_s(z_\ast,\bm{\theta})$ is the comoving sound horizon defined in Eq. \eqref{rs_general}, evaluated at $z_\ast$.

The redshift of photon decoupling, $z_\ast$, is obtained from the fitting formula \cite{Hu:1995en,Eisenstein:1997ik}

\begin{equation}
    z_\ast = 1048\left[1+0.00124\left(\Omega_{b0}h^2\right)^{-0.738}\right]
    \left[1+g_1\left(\Omega_{m0}h^2\right)^{g_2}\right],
\end{equation}
with

\begin{align}
    &g_1 = \frac{0.0783\left(\Omega_{b0}h^2\right)^{-0.238}}{1+39.5\left(\Omega_{b0}h^2\right)^{0.763}}, \\
    &g_2 = \frac{0.560}{1+21.1\left(\Omega_{b0}h^2\right)^{1.81}}.
\end{align}

The CMB chi-squared function is then defined as

\begin{equation}\label{chisqCMB}
    \chi^2_{\text{CMB}}= \bm{\Delta \textbf{X}^T_{\textrm{CMB}}\cdot \C^{-1}_{\textrm{CMB}}\cdot\Delta \textbf{X}_{\textrm{CMB}}},
\end{equation}
where $\bm{\Delta \textbf{X}_{\textrm{CMB}}} = \bm{\textbf{x}_{\textrm{CMB}}^{\textrm{obs}}}-\bm{\textbf{x}_{\textrm{CMB}}^{\textrm{th}}}$
is the difference between observed and theoretical values of the CMB distance priors
$\bm{\textbf{x}_{\textrm{CMB}}} = \{R,\,\ell_a,\,\Omega_{b0} h^2\}$,
and $\bm{\C^{-1}_{\textrm{CMB}}}$ is the inverse covariance matrix provided by \cite{Zhai:2018vmm}.

\subsection{Information Criteria}

To evaluate the relative effectiveness of each $f(Q)$ model in fitting the observational data, we adopt a statistical approach based on information-theoretic criteria. In particular, we compute the Akaike Information Criterion (AIC) \cite{Akaike:1974vps}, which provides a way to balance the efficiency of the fit against model complexity. For small sample sizes, the corrected $\textrm{AIC}_\textrm{C}$ is defined as \cite{Kenneth:2004a,Kenneth:2004b}
\begin{equation}\label{aic}
\textrm{AIC}_\textrm{C} = -2\ln{L_{\textrm{max}}}+2\kappa+\frac{2\kappa^2+2\kappa}{N-\kappa-1},
\end{equation}
where $L_{\textrm{max}}$ is the maximum likelihood achieved by the model given the data, $N$ denotes the total number of data points used in the fit, and $k$ is the number of free parameters in the model. Note that, in the limit of large $N$, the last term can be dropped out and the last expression reduces to the usual $\textrm{AIC}=-2\ln{L_{\textrm{max}}}+2\kappa$. Therefore, if the total number of data points is large enough, the use of the standard (reduced) AIC is sufficient \cite{Liddle:2007fy}. To compare each model with respect to $\Lambda$CDM, we calculate the AIC difference: $\Delta\textrm{AIC}_\textrm{C}(\Delta\textrm{AIC})\equiv\textrm{AIC}_{\textrm{C},f(Q)}(\textrm{AIC}_{f(Q)})-\textrm{AIC}_{\textrm{C},\Lambda\textrm{CDM}}(\textrm{AIC}_{\Lambda\textrm{CDM}})$. Since the total number of data points $N$ and the number of free parameters $k$ are the same for all the models considered (including $f(Q)$ and $\Lambda$CDM), the AIC comparison strongly simplifies to
\begin{equation}\label{aicdifference}
\Delta\textrm{AIC}_\textrm{C}(\Delta\textrm{AIC}) = -2\left[\ln{L^{f(Q)}_{\textrm{max}}}-\ln{L^{\Lambda\textrm{CDM}}_{\textrm{max}}}\right].
\end{equation}
In fact, as we see in this case  both $\Delta\textrm{AIC}_\textrm{C}$ and $\Delta\textrm{AIC}$ coincide. As a result, model selection in this context becomes straightforward: the model with the lower $\chi^2_{\textrm{min}}$ is preferred. To interpret the $\Delta\textrm{AIC}$ results, we follow the Jeffreys’ scale \cite{Jeffreys:1961a}, summarised in Tab. \ref{tab:jeffreys}.

\begin{table}[t]
    \centering
    \begin{tabular}{@{\hskip 15pt}c@{\hskip 15pt}c@{\hskip 15pt}}
        \hline
        \hline
        \addlinespace[2pt]
        $\Delta\textrm{AIC}$ & Interpretation \\
        \addlinespace[2pt]
        \hline
        \hline
        \addlinespace[2pt]
        $>10$  & Desively disfavoured \\
        \addlinespace[2pt]
        \hline
        \addlinespace[2pt]
        $5\sim10$  & Strongly disfavoured \\
        \addlinespace[2pt]
        \hline
        \addlinespace[2pt]
        $2\sim5$  & Moderately disfavoured \\
        \addlinespace[2pt]
        \hline
        \addlinespace[2pt]
        $-2\sim2$  & Compatible \\
        \addlinespace[2pt]
        \hline
        \addlinespace[2pt]
        $-5\sim-2$  & Moderately favoured \\
        \addlinespace[2pt]
        \hline
        \addlinespace[2pt]
        $-10\sim-5$  & Strongly favoured \\
        \addlinespace[2pt]
        \hline
        \addlinespace[2pt]
        $<-10$  & Decisively favoured \\
        \addlinespace[2pt]
        \hline
        \hline
    \end{tabular}
    \caption{\justifying{Interpretation of $\Delta\textrm{AIC}$ values based on     Jeffreys' scale. The scale provides qualitative guidance for comparing models relative to a reference model (here  $\Lambda$CDM), with negative values indicating preference for the alternative model and positive values favouring $\Lambda$CDM.}}
    \label{tab:jeffreys}
\end{table}

Another commonly used model selection tool is the Bayesian Information Criterion (BIC) \cite{Schwarz:1978tpv}, which incorporates a different penalisation for model complexity. The BIC is defined as \cite{Kenneth:2004a,Kenneth:2004b,Liddle:2007fy}
\begin{equation}\label{bic}
\textrm{BIC} = -2\ln{L_{\textrm{max}}}+\kappa\ln{N}.
\end{equation}
While AIC tends to favour models with better fit even if they are more complex, BIC introduces a stronger penalty for additional parameters, particularly for large datasets. As such, it provides a more conservative assessment of model preference. In our case, however, all the models under consideration are constrained using the same datasets and have the same number of free parameters. Consequently, the penalty term $\kappa\ln{N}$ is identical for all models and cancels out when computing the difference 
\begin{equation}\label{bicdifference}
\Delta\text{BIC} \equiv \text{BIC}_{f(Q)} - \text{BIC}_{\Lambda\text{CDM}} = \Delta\text{AIC}.
\end{equation}
This implies that the conclusions drawn from BIC are fully consistent with those obtained using AIC. Therefore, either criterion can be used for model comparison in our analysis without loss of generality.

\subsection{Results}

We now present the results of our parameter estimation analysis for the $f(Q)$ models, and we compare them with the standard $\Lambda$CDM cosmology. The analysis is performed for three different combinations of datasets: 

\begin{itemize}

\item \textbf{Combination I:} Cosmic chronometers (CC), supernovae (SN), and gamma-ray bursts (GRB);

\item \textbf{Combination II:} Baryon acoustic oscillations (BAO);

\item \textbf{Combination III:} Cosmic microwave background (CMB);

\item \textbf{Combination IV:} Baryon acoustic oscillations and cosmic microwave background (BAO + CMB);

\item \textbf{Combination V:} Full combination (CC + SN + GRB + BAO + CMB).

\end{itemize}

\begin{table*}[t]
    \centering
    \begin{tabular}{@{\hskip 15pt}c@{\hskip 15pt}c@{\hskip 15pt}c@{\hskip 15pt}c@{\hskip 15pt}c@{\hskip 15pt}}
        \hline
        \hline
        \addlinespace[2pt]
        Model & $H_0$ & $\Omega_{m0}$ & $\Omega_{b0}$ & $\Delta \text{AIC}$ \\
        \addlinespace[2pt]
        \hline
        \hline
        \addlinespace[2pt]
        \multicolumn{5}{c}{\textbf{CC + SN + GRB}} \\
        \addlinespace[2pt]
        $\Lambda$CDM  & $68.67 \pm 1.80$ & $0.3037 \pm 0.0202$ & $-$ & $-$ \\
        $f_1(Q)$  & $68.56 \pm 1.90$ & $0.3489 \pm 0.0206$ & $-$ & $0.192$ \\
        $f_2(Q)$  & $69.34 \pm 1.82$ & $0.2608 \pm 0.0159$ & $-$ & $1.65$ \\
        $f_3(Q)$  & $68.70 \pm 1.87$ & $0.3524 \pm 0.0226$ & $-$ & $-0.227$ \\
        \addlinespace[2pt]
        \hline
        \addlinespace[2pt]
        \multicolumn{5}{c}{\textbf{BAO}} \\
        \addlinespace[2pt]
        $\Lambda$CDM  & $73.2 \pm 15.0$ & $0.2948 \pm 0.0141$ & $0.0545 \pm 0.0160$ & $-$ \\
        $f_1(Q)$  & $74.4 \pm 14.7$ & $0.2897 \pm 0.0138$ & $0.0462 \pm 0.0148$ & $2.07$ \\
        $f_2(Q)$  & $70.6 \pm 14.9$ & $0.3036 \pm 0.0171$ & $0.0617 \pm 0.0180$ & $0.454$ \\
        $f_3(Q)$  & $73.6 \pm 13.8$ & $0.3065 \pm 0.0140$ & $0.0447 \pm 0.0130$ & $1.36$ \\
        \addlinespace[2pt]
        \hline
        \addlinespace[2pt]
        \multicolumn{5}{c}{\textbf{CMB}} \\
        \addlinespace[2pt]
        $\Lambda$CDM  & $67.273 \pm 0.633$ & $0.31649 \pm 0.00867$ & $0.049374 \pm 0.000692$ & $-$ \\
        $f_1(Q)$  & $71.821 \pm 0.709$ & $0.27794 \pm 0.00785$ & $0.043329 \pm 0.000664$ & $0.00249$ \\
        $f_2(Q)$  & $62.664 \pm 0.622$ & $0.36418 \pm 0.00982$ & $0.057004 \pm 0.000910$ & $-0.00365$ \\
        $f_3(Q)$  & $73.579 \pm 0.620$ & $0.26422 \pm 0.00641$ & $0.041322 \pm 0.000536$ & $-0.000683$ \\
        \addlinespace[2pt]
        \hline
        \addlinespace[2pt]
        \multicolumn{5}{c}{\textbf{BAO + CMB}} \\
        \addlinespace[2pt]
        $\Lambda$CDM  & $66.910 \pm 0.405$ & $0.32157 \pm 0.00562$ & $0.049831 \pm 0.000467$ & $-$ \\
        $f_1(Q)$  & $70.654 \pm 0.451$ & $0.29129 \pm 0.00526$ & $0.044473 \pm 0.000452$ & $3.74$ \\
        $f_2(Q)$  & $62.746 \pm 0.464$ & $0.36277 \pm 0.00730$ & $0.056875 \pm 0.000709$ & $6.01$ \\
        $f_3(Q)$  & $71.467 \pm 0.464$ & $0.28749 \pm 0.00534$ & $0.043213 \pm 0.000438$ & $16.0$ \\
        \addlinespace[2pt]
        \hline
        \addlinespace[2pt]
        \multicolumn{5}{c}{\textbf{CC + SN + GRB + BAO + CMB}} \\
        \addlinespace[2pt]
        $\Lambda$CDM  & $67.022 \pm 0.377$ & $0.32001 \pm 0.00523$ & $0.049719 \pm 0.000443$ & $-$ \\
        $f_1(Q)$  & $70.431 \pm 0.460$ & $0.29392 \pm 0.00543$ & $0.044673 \pm 0.000465$ & $11.0$ \\
        $f_2(Q)$  & $63.621 \pm 0.419$ & $0.34929 \pm 0.00627$ & $0.055671 \pm 0.000603$ & $37.1$ \\
        $f_3(Q)$  & $70.942 \pm 0.254$ & $0.29362 \pm 0.00295$ & $0.043691 \pm 0.000285$ & $23.2$ \\
        \addlinespace[2pt]
        \hline
        \hline
    \end{tabular}
    \caption{\justifying{Mean values and standard deviations of the cosmological parameters obtained for each $f(Q)$ model, namely $ f_1(Q) = Q\,e^{\lambda_1 Q_0/Q}$, $ f_2(Q) = Q+Q_0 e^{-\lambda_2 Q_0/Q}$, and $f_3(Q) = Q+\lambda_3 Q_0\!\left[1-e^{-Q_0/Q}\right]$, 
  and for $\Lambda$CDM paradigm, under the five different dataset combinations considered in this work: \textbf{Combination I} (CC + SN + GRB), \textbf{Combination II} (BAO), \textbf{Combination III} (CMB), \textbf{Combination IV} (BAO + CMB), and \textbf{Combination V} (CC + SN + GRB + BAO + CMB). The last column shows the AIC difference, $\Delta \text{AIC} \equiv \text{AIC}_{f(Q)} - \text{AIC}_{\Lambda\text{CDM}}$, quantifying the statistical preference relative to the $\Lambda$CDM model.}}
    \label{tab:fq_models}
\end{table*}

Table \ref{tab:fq_models} summarises the mean values and standard deviations of the cosmological parameters. As stated at the beginning of this section, the baryon density $\Omega_{b0}$ is treated as a free parameter only in Combinations~II$-$V, whereas Combination~I depends solely on $H_0$ and $\Omega_{m0}$.

Under Combination I, which includes only background cosmological probes (CC, SN, and GRB), two of the $f(Q)$ models (specifically Model 1 and Model 3) favour higher values of the matter density parameter $\Omega_{m0}$ compared to the $\Lambda$CDM scenario. Conversely, Model 2 prefers a lower matter density. The Hubble constant $H_0$ is similarly constrained across all models, exhibiting very close central values and uncertainties. This behaviour is consistently reflected in the values reported in Table \ref{tab:fq_models}.

Under Combination II, which includes only baryon acoustic oscillation (BAO) data, the Hubble constant $H_0$ and the baryon density parameter $\Omega_{b0}$ are weakly constrained, leading to large uncertainties as seen in Table \ref{tab:fq_models}. In contrast, the matter density parameter $\Omega_{m0}$ is relatively well determined and shows consistent values across all $f(Q)$ models and the $\Lambda$CDM scenario. As a consequence, BAO data alone do not have the constraining power to significantly impact the discussion of the $H_0$ tension, and their main contribution emerges only when combined with early-time probes such as the CMB.

Under Combination III, which includes only cosmic microwave background (CMB) data, all cosmological parameters are very tightly constrained, as expected from the strong constraining power of early-universe observations. In this case, the three $f(Q)$ models exhibit values of $H_0$, $\Omega_{m0}$, and $\Omega_{b0}$ that deviate systematically from the $\Lambda$CDM predictions: Models 1 and 3 favour larger $H_0$ and smaller $\Omega_{m0}$ and $\Omega_{b0}$, while Model 2 predicts the opposite trend, with a lower $H_0$ and higher matter and baryon densities. Importantly, Models 1 and 3 yield $H_0$ estimates closer to the local measurements of \cite{Riess:2020fzl}, thereby alleviating the $H_0$ tension, whereas Model 2 exacerbates it by driving $H_0$ to even smaller values than in $\Lambda$CDM.

Under Combination IV, which combines baryon acoustic oscillation (BAO) and cosmic microwave background (CMB) data, the cosmological parameters are constrained with very high precision. Models 1 and 3 predict higher values of the Hubble constant $H_0$ than $\Lambda$CDM, accompanied by lower values of $\Omega_{m0}$ and $\Omega_{b0}$, bringing their $H_0$ estimates closer to the local measurements of \cite{Riess:2020fzl} and thereby partially alleviating the $H_0$ tension. In contrast, Model 2 yields a smaller $H_0$ relative to $\Lambda$CDM, together with higher $\Omega_{m0}$ and $\Omega_{b0}$, which aggravates the Hubble tension. This complementary behaviour between Models 1/3 and Model 2 mirrors the trends already observed in the CMB-only analysis.

\begin{figure*}[t]  
    \centering
    \includegraphics[width=0.9\textwidth]{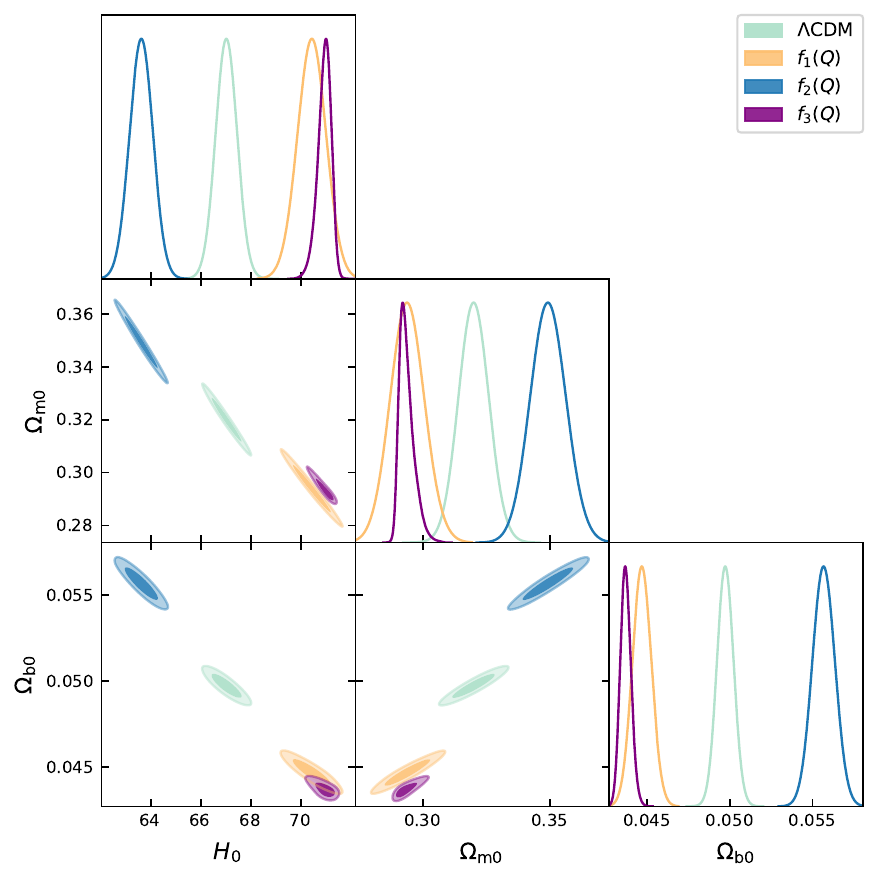}  
    \caption{\justifying{{\it{Two-dimensional posterior distributions for the $f(Q)$ models and $\Lambda$CDM scenario, using Combination V (CC + SN + GRB + BAO + CMB). The contours correspond to the 68\% and 95\% confidence levels (C.L.). This figure summarises the full parameter constraints, including $\Omega_{b0}$. It illustrates how Models 1 and 3 accommodate higher values of $H_0$ in contrast to Model 2, which yields a lower $H_0$ compared to $\Lambda$CDM.}}}}
    \label{fig:cc_sn_grb_bao_cmb}
\end{figure*}

Under Combination V, where all cosmological datasets are combined (CC + SN + GRB + BAO + CMB), Models 1 and 3 continue to allow for higher values of the Hubble constant $H_0$ relative to $\Lambda$CDM, with estimates lying closer to direct local measurements \cite{Riess:2020fzl}. At the same time, both models predict lower values of $\Omega_{m0}$ and $\Omega_{b0}$. In contrast, Model 2 yields a smaller value of $H_0$ compared to $\Lambda$CDM, thereby exacerbating the Hubble tension, while simultaneously favouring larger values of $\Omega_{m0}$ and $\Omega_{b0}$. These behaviours are clearly illustrated in Fig.~\ref{fig:cc_sn_grb_bao_cmb}, where a strong correlation between the parameters is observed: higher values of $H_0$ correspond to lower values of both $\Omega_{m0}$ and $\Omega_{b0}$, and vice versa.

\begin{figure*}[t]
    \centering
    \begin{tabular}{cc}
        \includegraphics[width=0.5\textwidth]{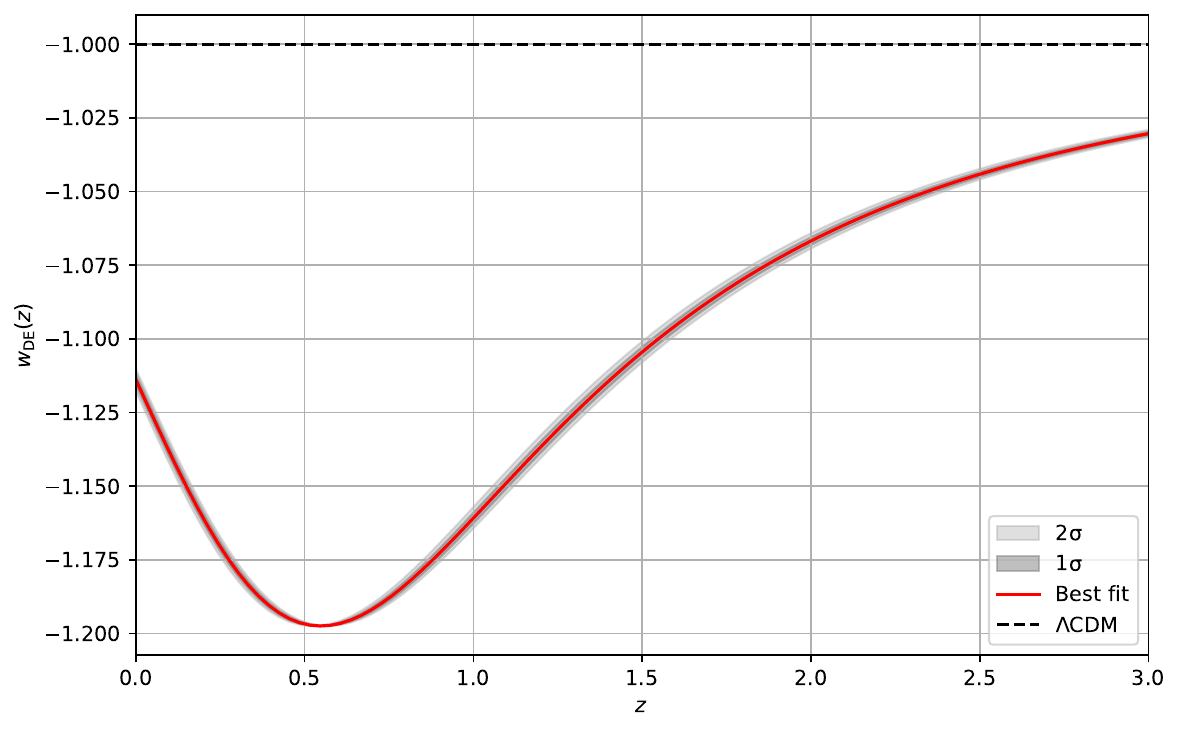} &
        \includegraphics[width=0.5\textwidth]{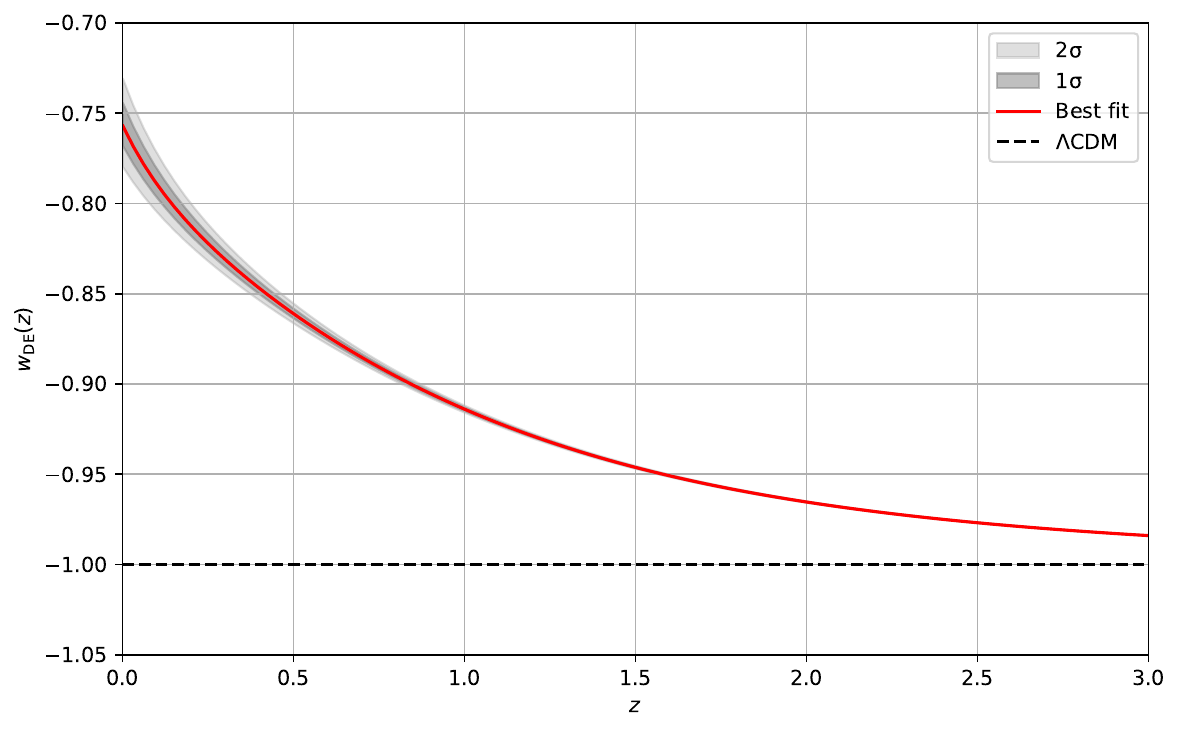} \\
        \multicolumn{2}{c}{\includegraphics[width=0.6\textwidth]{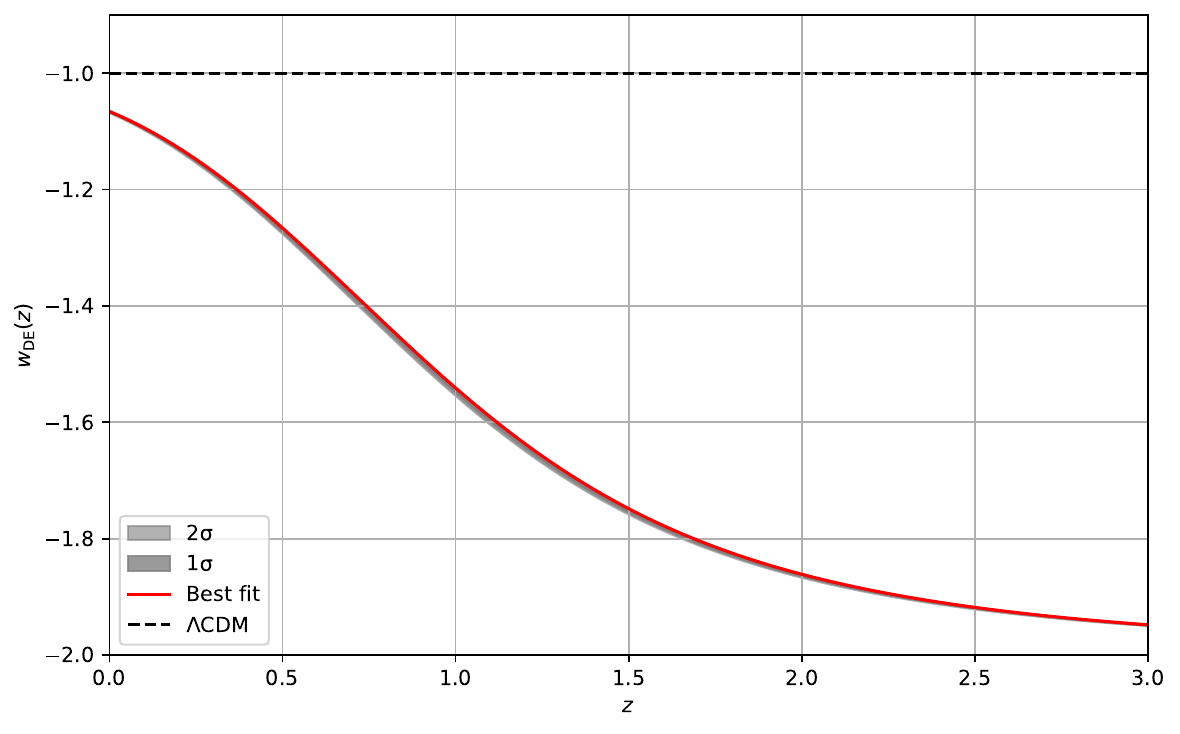}} \\
    \end{tabular}
    \caption{\justifying{{\it{
    Reconstruction of the effective dark energy equation of state $w_{\mathrm{DE}}(z)$ 
    for the three $f(Q)$ models using the full dataset combination (Combination~V). 
    Solid curves correspond to the best-fit reconstruction, 
    while shaded regions represent the 68\% (dark) and 95\% (light) confidence levels 
    obtained from the MCMC analysis. 
    The three panels show: Model~1 (top-left), Model~2 (top-right), and Model~3 (bottom). 
    Compared to Fig.~\ref{fig:theory_motivation}, which displayed only the best-fit behaviours, 
    here we explicitly include the confidence regions, demonstrating that the constraints on 
    $w_{\mathrm{DE}}$ are consistently tight across all three models.}}}}
    \label{fig:wde_recon}
\end{figure*}

Before concluding, we briefly comment on the implications of our results for the so-called $S_8$ tension. In our statistical analysis we did not include $S_8$ as a fitted parameter. Nevertheless, some qualitative insights can be drawn from the theoretical behaviour of the effective gravitational coupling $G_{\mathrm{eff}}$ and the effective dark energy equation of state $w_{\mathrm{DE}}$ in the three $f(Q)$ models (see Fig.~\ref{fig:theory_motivation}). In particular, only Model~2 satisfies $G_{\mathrm{eff}}<G$ (bottom panel), a feature usually associated with a suppression of structure growth and therefore with the potential to alleviate the $S_8$ tension. On the other hand, Models~1 and~3 exhibit $G_{\mathrm{eff}}>G$ at all redshift, which typically enhances clustering and may worsen the tension. Similar conclusions have been discussed in the literature \cite{Heisenberg:2022lob,Heisenberg:2022gqk}, where it has been noted that scenarios with $G_{\mathrm{eff}}>G$ tend to aggravate the $S_8$ problem. Regarding the effective dark energy dynamics, Models~1 and~3 display phantom-like behaviour ($w_{\mathrm{DE}}<-1$), while Model~2 behaves in a quintessence-like manner ($w_{\mathrm{DE}}>-1$), as shown in the top-right panel of Fig.~\ref{fig:theory_motivation}. Although this correspondence is not a universal feature of all modified gravity or dark energy models, it has been observed in several cases that phantom-like scenarios, which can ease the $H_0$ tension, often worsen the $S_8$ tension, while quintessence-like scenarios behave in the opposite way. The $f(Q)$ models analysed here appear to follow this trend: Models~1 and~3 bring $H_0$ closer to local measurements but aggravate $S_8$, whereas Model~2 has the potential to mitigate $S_8$ but exacerbates the $H_0$ discrepancy. This complementarity highlights the difficulty of simultaneously addressing both tensions within the minimal $f(Q)$ scenarios considered here. For completeness, we also present in Fig.~\ref{fig:wde_recon} the full statistical reconstruction of $w_{\mathrm{DE}}(z)$, obtained from the MCMC chains using Combination~V. These plots extend the illustrative best-fit curves of Fig.~\ref{fig:theory_motivation} by explicitly including the 68\% and 95\% confidence regions. The narrowness of the reconstructed bands confirms that $w_{\mathrm{DE}}$ is tightly constrained, with only mild deviations from the best-fit behaviour allowed. This reinforces the qualitative conclusions drawn above and further highlights the limited freedom available to $f(Q)$ models in the effective dark energy sector.

We conclude this section by comparing the three $f(Q)$ models against the standard $\Lambda$CDM cosmology using the Akaike Information Criterion (AIC), as described in the previous subsection. According to the Jeffreys’ scale (Table \ref{tab:jeffreys}), we find that:

\begin{itemize}

\item Under Combination I, all $f(Q)$ models are statistically compatible with $\Lambda$CDM, with $\Delta\mathrm{AIC}$ values indicating no significant preference for either model class.

\item For Combination II (BAO alone), all $f(Q)$ models are statistically compatible with $\Lambda$CDM. Model~1 is slightly disfavoured ($\Delta\mathrm{AIC}\!\approx\!2$), while Models~2 and~3 show no significant preference relative to $\Lambda$CDM.

\item For Combination III (CMB alone), all $f(Q)$ models are statistically compatible with $\Lambda$CDM, with no significant preference for either model class.

\item In contrast, under Combination IV (BAO + CMB), all $f(Q)$ models are in tension with $\Lambda$CDM, with $\Lambda$CDM being clearly favoured.

\item When considering Combination V, which includes all cosmological probes, the tension with $\Lambda$CDM is further exacerbated, and $\Lambda$CDM is strongly favoured.
    
\end{itemize}

\begin{figure*}[t]  
    \centering
    \begin{tabular}{cc}  
        \includegraphics[width=0.45\linewidth]{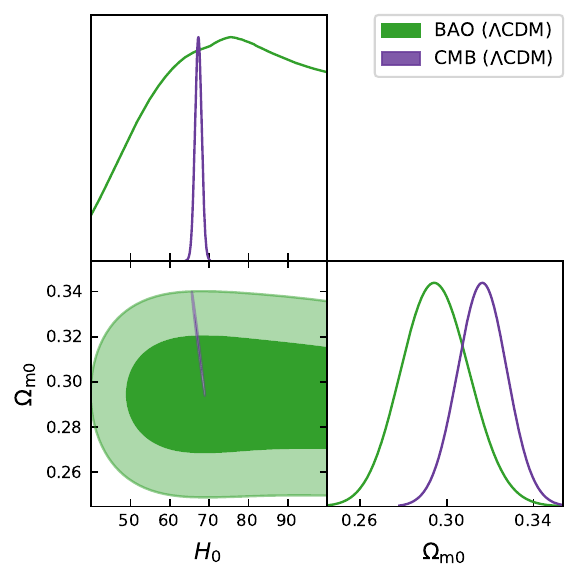} &
        \includegraphics[width=0.45\linewidth]{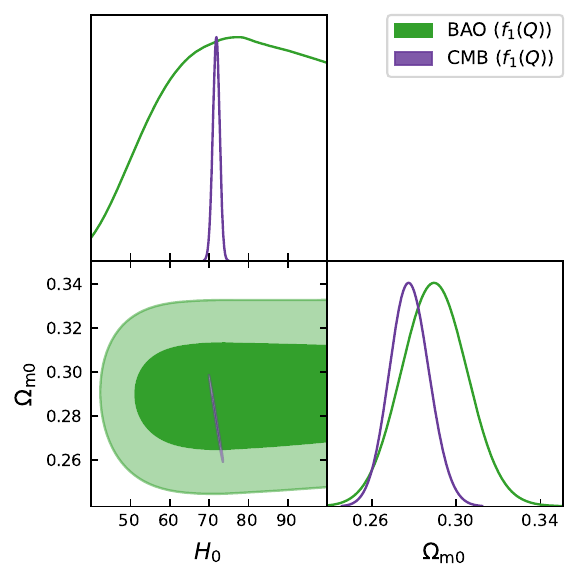} \\
        \includegraphics[width=0.45\linewidth]{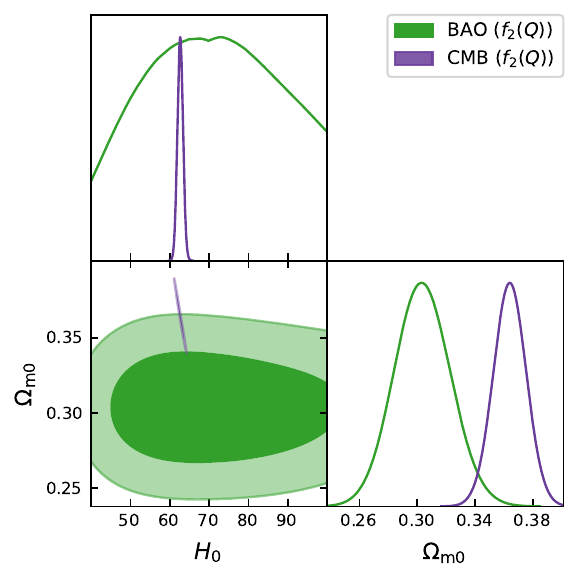} &
        \includegraphics[width=0.45\linewidth]{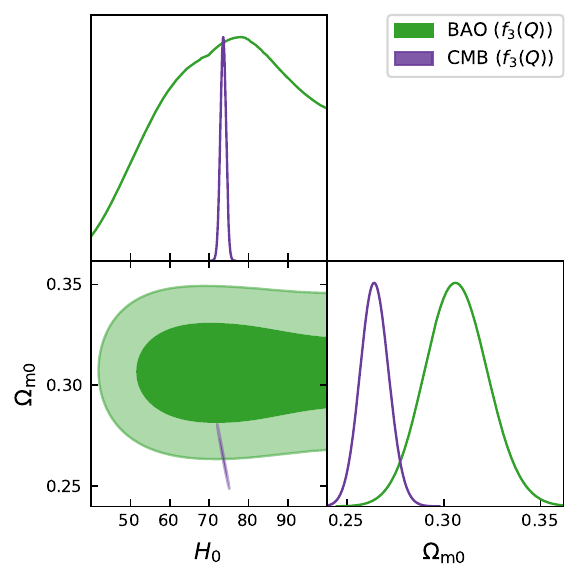} \\
    \end{tabular}
    \caption{\justifying{{\it{Comparison of the two-dimensional posterior distributions obtained from Combination II (BAO) and Combination III (CMB) for each model separately. The contours correspond to the 68\% and 95\% confidence levels (C.L.). The top-left panel shows the results for $\Lambda$CDM scenario, which displays excellent agreement between BAO and CMB. The remaining panels correspond to Models 1 (top-right), 2 (bottom-left), and 3 (bottom-right). A clear tension between BAO and CMB in the $\Omega_{m0}-H_0$ plane appears in Models 2 and 3.}}}}
    \label{fig:tensionsbaocmb}
\end{figure*}

The above results indicate that, in the BAO + CMB case (Combination IV), $\Lambda$CDM is clearly preferred over the $f(Q)$ models. Model~1 is only mildly disfavoured, consistent with the trend already observed when considering BAO data alone, but it still shows good agreement between BAO and CMB datasets. By contrast, Models~2 and~3 exhibit the strongest internal mismatch between BAO and CMB constraints. This behaviour is clearly visible in Fig.~\ref{fig:tensionsbaocmb}, which displays the joint posterior distributions at $68\%$ and $95\%$ confidence levels (C.L.) for each model, comparing the constraints from Combination~II (BAO) and Combination~III (CMB). While $\Lambda$CDM and Model~1 show excellent agreement between the two datasets, Models~2 and~3 reveal a clear tension in the $\Omega_{m0}-H_0$ plane, which directly translates into their poorer statistical performance and disfavour with respect to $\Lambda$CDM under Combination~IV.

\begin{figure*}[t]  
    \centering
    \begin{tabular}{cc}  
        \includegraphics[width=0.45\linewidth]{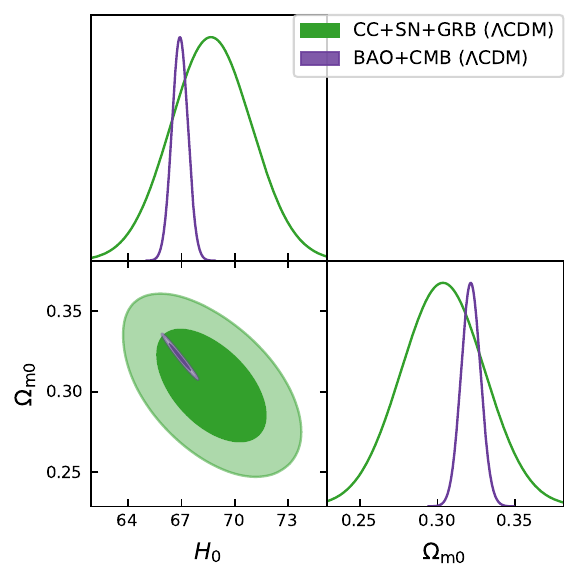} &
        \includegraphics[width=0.45\linewidth]{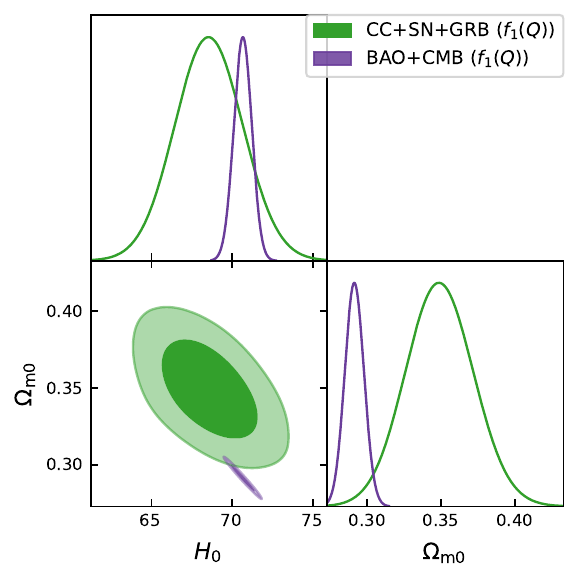} \\
        \includegraphics[width=0.45\linewidth]{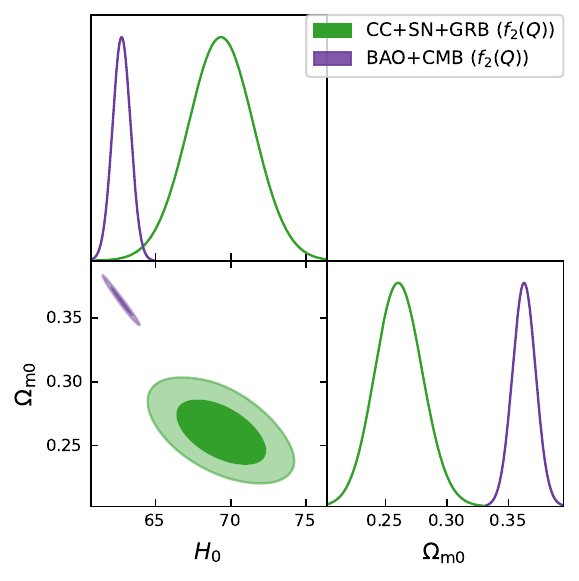} &
        \includegraphics[width=0.45\linewidth]{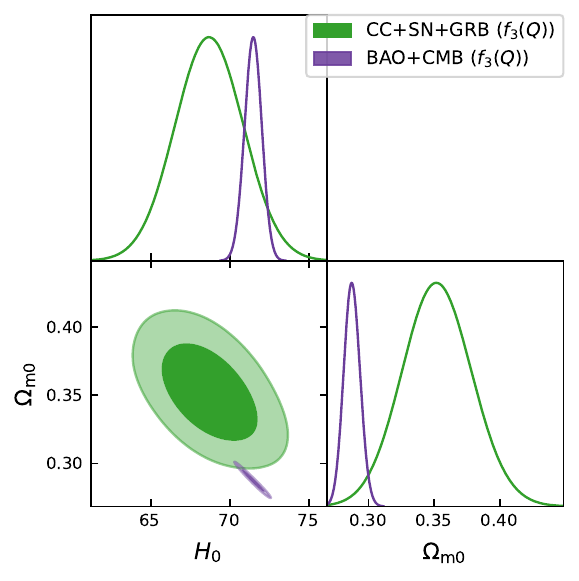} \\
    \end{tabular}
    \caption{\justifying{{\it{Comparison of the two-dimensional posterior distributions obtained from Combination I (CC + SN + GRB) and Combination IV (BAO + CMB) for each model separately. The contours correspond to the 68\% and 95\% confidence levels (C.L.). The top-left panel shows the results for $\Lambda$CDM scenario, which displays excellent agreement between the dataset combinations. The remaining panels correspond to Models 1 (top-right), 2 (bottom-left), and 3 (bottom-right), where a clear tension between the combinations emerges in the $\Omega_{m0}-H_0$ plane. These internal inconsistencies contribute to the poorer global fits obtained by the $f(Q)$ models when all datasets are combined.}}}}
    \label{fig:tensionsfull}
\end{figure*}

The comparison between Combinations~IV (BAO + CMB) and~V (all datasets combined) highlights the internal tensions within the $f(Q)$ framework when background information is confronted with early-time and large-scale structure probes. As already noted, background-only data (Combination~I) push Models~1 and~3 toward higher values of $\Omega_{m0}$ relative to $\Lambda$CDM, while Model~2 prefers a lower matter density. In contrast, when BAO and CMB are combined (Combination~IV), all three $f(Q)$ models shift in the opposite direction, thereby generating clear inconsistencies between dataset combinations. These mismatches are clearly visible in Fig.~\ref{fig:tensionsfull}, which displays the joint posterior distributions at $68\%$ and $95\%$ confidence levels (C.L.) for each model, comparing Combination~I (CC + SN + GRB) with Combination~IV (BAO + CMB). While the $\Lambda$CDM case shows excellent agreement between the datasets, all three $f(Q)$ models exhibit visible tensions in the $\Omega_{m0}-H_0$ plane. Among them, Model~2 displays the strongest discrepancy, consistent with its poorest statistical performance under the AIC analysis. These tensions, already present in Combination~IV, propagate into the full Combination~V, ultimately reducing the efficiency of the global fits for the $f(Q)$ scenarios relative to $\Lambda$CDM.

We note that similar internal tensions have been reported in related $f(Q)$ models, as discussed in \cite{Ayuso:2020dcu}. That analysis considered both cases, with and without a cosmological constant term. Interestingly, the mismatch between different dataset combinations arises specifically in the case $\Omega_\Lambda = 0$, whereas models with a non-vanishing $\Omega_\Lambda$ yield consistent fits. This suggests that the origin of the tension may be intrinsically tied to the absence of a cosmological constant, pointing to a potential limitation of purely geometric $f(Q)$ models without $\Lambda$. Further theoretical and phenomenological work will be required to clarify whether extended formulations of the theory can overcome this issue.

\section{Conclusions}
\label{conclusions}

In this work we have investigated the potential of modified gravity theories based on non-metricity, specifically $f(Q)$ gravity, to address two of the most persistent cosmological tensions: the discrepancy in the Hubble constant $H_0$ and the $S_8$ tension related to the growth of structure. We considered three representative $f(Q)$ models and performed a comprehensive parameter estimation analysis using a wide range of cosmological observations, combined in five distinct dataset configurations: (I) CC + SN + GRB, (II) BAO, (III) CMB, (IV) BAO + CMB, and (V) the full combination of all probes.

A key result of our analysis is that $f(Q)$ models can shift cosmological predictions in directions relevant to both tensions. In particular, Models~1 and~3 predict higher values of the Hubble constant $H_0$ compared to $\Lambda$CDM when CMB data are included (Combination~III), bringing the estimates closer to direct local determinations and partially relieving the $H_0$ tension. Model~2, on the other hand, exhibits $G_{\mathrm{eff}}<G$, a property typically associated with suppressed growth of structure and therefore with the potential to ease the $S_8$ tension. The theoretical behaviour of these models, illustrated in Fig.~\ref{fig:theory_motivation}, helps interpret these findings: Models~1 and~3 display phantom-like effective dark energy dynamics ($w_{\mathrm{DE}}<-1$) together with $G_{\mathrm{eff}}>G$, while Model~2 follows a quintessence-like trajectory ($w_{\mathrm{DE}}>-1$) with $G_{\mathrm{eff}}<G$. This variety of late-time behaviours is consistent with the general trend observed in many modified gravity and dark energy scenarios, where models that raise $H_0$ often worsen $S_8$, while those that suppress clustering tend to prefer lower $H_0$. These qualitative findings are further supported by the full reconstruction of $w_{\mathrm{DE}}(z)$, shown in Fig.~\ref{fig:wde_recon}, where the narrowness of the $68\%$ and $95\%$ confidence regions confirms that deviations from the best-fit trajectories are strongly constrained.

At the same time, our analysis uncovered certain internal tensions within the $f(Q)$ framework when different dataset combinations are confronted. While background probes (Combination~I) favour larger values of $\Omega_{m0}$ in Models~1 and~3, the BAO and CMB datasets (Combinations~II and III) drive them toward lower values. Model~2 shows the opposite trend, preferring lower $\Omega_{m0}$ with background data but higher values with BAO and CMB. These mismatches, clearly illustrated in Fig.~\ref{fig:tensionsfull}, reveal inconsistencies between the background constraints and those from early-time and large-scale structure formation. Nevertheless, it is important to stress that all three models remain statistically viable for individual dataset combinations, and only when probes are combined does $\Lambda$CDM emerge as the more efficient global description. Similar issues have been reported in related studies \cite{Ayuso:2020dcu}, where introducing a cosmological constant improves the consistency of the fits. This suggests that the absence of $\Lambda$ in the minimal $f(Q)$ formulations studied here may be at the origin of these residual tensions.

In summary, $f(Q)$ gravity offers a theoretically motivated framework capable of alleviating either the $H_0$ or the $S_8$ tension, depending on the model considered. Models~1 and~3 are favoured in relation to $H_0$, while Model~2 is favoured in relation to $S_8$. Future work should therefore explore extended $f(Q)$ formulations, such as including a cosmological constant, adopting more general functional forms, or analysing other connection choices. In addition, testing these models against further observational probes, such as weak lensing and galaxy clustering, will be crucial to fully assess their viability. 

We finally remark that the variations of $G_{\mathrm{eff}}$ discussed here occur 
on cosmological timescales and converge to $G$ at early-time ($f_Q \to 1$), thus 
preserving agreement with well-tested early-universe physics. A detailed analysis 
of possible implications for local gravity constraints, where no well-established 
screening mechanism is currently known in $f(Q)$ gravity (unlike in scalar–tensor 
theories), lies beyond the scope of this work.

Continued investigation along these lines will help clarify whether non-metricity-based gravity can serve as a compelling alternative to the concordance cosmological model.
\vspace{-0.4cm}
\acknowledgments
CGB acknowledges financial support from the FPI fellowship PRE2021-100340 
of the Spanish Ministry of Science, Innovation and Universities. 
MB-L is supported by the Basque Foundation of Science Ikerbasque. Our work 
is supported by the Spanish grant PID2023-149016NB-I00 (funded by MCIN/AEI/10.13039/501100011033 and by “ERDF A way of making Europe"). This research is also supported by 
the Basque government Grant No. IT1628-22 (Spain).  
The authors acknowledge the contribution of the COST Action CA21136 “Addressing 
observational tensions in cosmology
with systematics and fundamental physics (CosmoVerse)”.

\bibliography{bibliografia}

\end{document}